\documentclass{aa}  

\usepackage{graphicx}
\usepackage{txfonts}
\newcommand{\flash}{\textsc{FLASH}}
\newcommand{\tevol}{t_\mathrm{evol}}
\newcommand{\tdust}{T_\mathrm{d}}
\newcommand{\deficit}{I_\mathrm{[CII]}/I_\mathrm{FIR}}
\newcommand{\deficitl}{L_\mathrm{[CII]}/L_\mathrm{FIR}}
\newcommand{\intunity}{erg s$^{-1}$ cm$^{-2}$ sr$^{-1}$}
\newcommand{\lstartot}{L_{\star \mathrm{,tot}}}

\begin{document}

\title{The origin and evolution of the [CII] deficit in HII regions and star-forming molecular clouds}

\titlerunning{The {[}CII{]}-deficit in HII regions and molecular clouds}

\author{
S. Ebagezio\inst{1}
\and
D. Seifried\inst{1,2}\fnmsep\thanks{E-mail: seifried@ph1.uni-koeln.de}
\and
S. Walch\inst{1,2}
\and
T.~G. Bisbas\inst{3,1}
}

\institute{
Universit\"at zu K\"oln, I. Physikalisches Institut, Z\"ulpicher Str.~77, 50937 K\"oln, Germany
\and
Center for Data and Simulation Science, University of Cologne, Germany
\and
Research Center for Astronomical Computing, Zhejiang Laboratory, Hangzhou 311100, China
}

\date{Received February 2024}

\abstract
{}
{We analyse synthetic emission maps of the [CII] 158 $\mu$m line and far-infrared (FIR) continuum of simulated molecular clouds (MCs) within the SILCC-Zoom project to study the origin of the observed [CII] deficit, that is, the drop in the [CII]/FIR intensity ratio caused by stellar activity.}
{All simulations include stellar radiative feedback and the on-the-fly chemical evolution of hydrogen species, CO, and C$^+$. We also account for further ionisation of C$^+$ into C$^{2+}$ inside HII regions, which is crucial to obtain reliable results.}
{
Studying individual HII regions, we show that $I_\mathrm{FIR}$ is initially high in the vicinity of newly born stars, and then moderately decreases over time as the gas is compressed into dense and cool shells.
In contrast, there is a large drop in $I_\mathrm{CII}$  over time, to which the second ionisation of C$^+$ into C$^{2+}$ contributes significantly. 
This leads to a large drop in $\deficit$ inside HII regions, with $\deficit$ decreasing from 10$^{-3}$--10$^{-2}$ at scales above 10~pc to around 10$^{-6}$--10$^{-4}$ at scales below 2~pc. However, projection effects can significantly affect the radial profile of $I_\mathrm{[CII]}$, $I_\mathrm{FIR}$, and their ratio, and can create apparent HII regions without any stars.
Considering the evolution on MC scales, we show that the luminosity ratio, $\deficitl$, decreases from values of $\gtrsim$10$^{-2}$ in MCs without star formation to values of around $\sim10^{-3}$ in MCs with star formation.
We attribute this decrease and thus the origin of the [CII] deficit to two main contributors: (i) the saturation of the [CII] line and (ii) the conversion of C$^+$ into C$^{2+}$ by stellar radiation.
The drop in the $\deficitl$ ratio can be divided into two phases: (i)
During the early evolution of HII regions, the saturation of [CII] and the further ionisation of C$^+$ limit the increase in $L_\mathrm{[CII]}$, while $L_\mathrm{FIR}$ increases rapidly, leading to the initial decline of $\deficitl$.
(ii) In more evolved HII regions, $L_\mathrm{CII}$ stagnates and even partially drops over time due to the aforementioned reasons.
$L_\mathrm{FIR}$ also stagnates as the gas gets pushed into the cooler shells surrounding the HII region. In combination, this keeps the global $\deficitl$ ratio at low values of $\sim10^{-3}$.}
{}


\maketitle

\section{Introduction}
\label{sec:introduction}

Molecular clouds (MCs) are the densest and coldest regions of the interstellar medium (ISM) and, as the name suggests, are defined as those regions where hydrogen exists mainly in molecular form. Such environments are also the regions where dense cores and stars form. The formation and evolution of MCs has been studied with numerical simulations in a large number of recent works \citep[see e.g.~the reviews by][and references therein]{Henshaw2023, Chevance2023,Hacar2023}. Once star formation begins, it is of particular interest to explore the effects that stellar feedback has on the evolution and possible disruption of MCs. Here, the three most important forms of stellar feedback are supernova explosions, stellar winds, and ionising radiation. In particular, the latter can contribute to cloud dispersal as it decreases the density of the cloud in and around sites of star formation. As a consequence, this also further inhibits star formation, contributing to the low star formation efficiency generally observed \citep[e.g.][]{Krumholz2014,Peters2017,Rathjen21,Rathjen23}. In addition, photons emitted by stars dissociate or ionise the surrounding gas. The first theoretical study of this phenomenon, performed by \citet{Stromgren1939}, showed that, under the assumption of uniform density and isotropic conditions, stellar radiation produces a shell where hydrogen is ionised; the radius $r$ of this shell is \mbox{$r \propto Q_0^{1/3} \, n^{-2/3}$}, where $Q_0$ is the ionising flux and $n$ the gas density. These portions of MCs are called HII regions. Later on, \citet{Spitzer1978} calculated the expansion velocity of the ionisation fronts, finding $r \propto t^{4/7}$ \citep[see also][]{Bisbas15,Williams18}. Following on from these fundamental works, ionising radiation and its role in MCs have been extensively studied in a large number of numerical works \citep[][and many more]{Hosokawa2006, Krumholz2009, Fall2010, Murray2010, Dale2012, Walch2012, Girichidis2016, Gatto2017, Haid2019, Scheuermann2023}.

The radiation from the stars has a substantial impact on the chemistry of the clouds. At the boundaries of the aforementioned HII regions, the so-called photodissociation regions (PDRs) form, where the chemistry is dominated by far-ultraviolet (FUV) photons and their progressive attenuation \citep[][and references therein]{Wolfire2022}. In this context, one of the brightest emission lines originating from star-forming regions is the [CII] line at 157.7 $\mu$m \citep[e.g.][]{Stacey1991, Brauher2008}, which is one of the main coolants in the warm ISM \citep{Tielens1985, Stacey1991, Hollenbach1999, Stutzki2001, Rollig2006, Wolfire2003, Ossenkopf2013, Beuther2014, Pineda2013, Pineda2014, Klessen2016}. The [CII] emission has been extensively observed towards HII regions \citep{Schneider2012, Goicoechea2015, Rollig2016, Pabst2017, Pabst2020, Schneider2020, Tiwari2021, Luisi2021, Beuther2022, Kabanovic2022}, and is found to generally highlight a bubble structure with bright rims and very little emission from the inner part. [CII] is also frequently used as a tracer of the star-formation rate in galaxies \citep[e.g.][]{Stacey1991,Boselli02,DeLooze14,Herrera2015,Sutter2019,Bisbas2022} as well as a tracer in cloud--cloud interactions \citep{Bisbas17,Schneider23}.

The far-infrared (FIR) dust continuum emission is the thermal emission from dust grains with temperatures in a typical range of \mbox{$10 - 50$ K}. 
It is usually defined as the integrated emission in the wavelength range from 40~$\mu$m to 500~$\mu$m \citep[e.g.][]{Sanders1996, Goicoechea2015, Pabst2021}.
It is also a meaningful measure for understanding HII regions, as it is intimately associated with star formation, as shown in numerous recent studies \citep[][]{Brauher2008, Gracia2011, Delooze2011, Casey2014, Herrera2018, Pabst2022, Dunne2022}. 

The ratio between [CII] and FIR emission is of particular interest for understanding stellar feedback in MCs as both are thought to trace star formation. However, extragalactic observations show that this ratio, which is typically of the order of \mbox{$10^{-2} - 10^{-3}$}, decreases with increasing infrared luminosity across galaxies \citep{Malhotra1997, Luhman1998, Malhotra2001, Luhman2003, Casey2014,Diaz2017, Smith2017, Herrera2018}. The origin of this so-called [CII] deficit is still unclear, although different hypotheses have been proposed, including AGN contributions to FIR emission \citep{Sargsyan2012}, further ionisation of C$^+$ into C$^{2+}$ \citep{Abel2005}, fine-structure lines overcoming [CII] as coolants \citep[][for instances]{Luhman2003}, and saturation of the [CII] line \citep{Munoz2016, Rybak2019, Bisbas2022}. Evidence for a relationship between the [CII] deficit and the star formation rate (SFR) has also been presented, both theoretical \citep{Narayanan2017} and observational \citep{Stacey2010, Smith2017, Hu2019}. To get a better understanding of the origin of the [CII] deficit, resolved studies of individual HII regions and MCs are required. However, so far only a few have been reported in the literature, and these explicitly consider both the [CII] and FIR emission on MC scales \citep{Goicoechea2015,Pabst2021}.

Here, we extend our previous work on synthetic emission maps from MCs \citep{Ebagezio2023} and provide a simulation counterpart to the two [CII] and FIR observations on MC scales mentioned above in order to enable a comparison between real and simulated data, and to take advantage of our knowledge of the simulation input to obtain more detailed insights into the physical phenomena that cause the [CII] deficit.
To this end, we used the SILCC-Zoom simulations, which include stellar feedback and a chemical network to properly account for the ionised carbon from which the [CII] emission emerges \citep{Seifried2017,Haid2019}. The paper is organised as follows. In Section \ref{sec:numerics} we present the numerical simulations of the MCs that serve as a basis for our analysis, as well as the post-processing and analysis techniques that we use. Next, in Section~\ref{sec:results} we describe our results: we investigate the dust temperature, and the [CII] and FIR emission maps and their ratio. These results are discussed in Section~\ref{sec:discussion}, where we present a more detailed investigation of the causes of the [CII] deficit, and make comparisons to observational data. In Section~\ref{sec:conclusions} we summarise our work and draw conclusions. 

\section{Numerical methods}\label{sec:numerics}

This work is a further analysis of the synthetic emission maps described in \citet{Ebagezio2023}. Therefore, we refer to this paper and to the related SILCC-Zoom simulation papers \citep{Seifried2017,Haid2019} for a detailed description of the numerical simulations, the post-processing routines, and the radiative transfer. Here, we briefly summarise the most important concepts of these simulations, and describe the additional processes not incorporated into the aforementioned works.

\subsection{SILCC-Zoom simulations}
\label{sec:silcc_zoom_simulations}

The simulated MCs that we use in this work are part of the SILCC-Zoom project \citep[][Walch et al., in prep.]{Seifried2017,Haid2019}. This consists of zoom-in simulations of specific regions of the larger-scale simulations within the SILCC project \citep[see][for details]{Walch2015, Girichidis2016}. 
    
The SILCC project aims to simulate the life-cycle of molecular clouds in typical solar neighbourhood conditions. To do so, it models a region of a stratified galactic disc of 500~pc $\times$ 500~pc $\times$ $\pm$5~kpc in size using the adaptive mesh refinement code \textsc{FLASH 4.3} \citep{Fryxell2000, Dubey2008}. The maximum allowed resolution is 3.9~pc. Both hydrodynamical (HD) and magnetohydrodynamical (MHD) runs are simulated within the project. The chemistry is modelled using a chemical network for H$^+$, H, H$_2$, C$^+$, O, CO, and e$^-$ \citep{Nelson1997, Glover2007a, Glover2007b, Glover2010}, also including the most important heating and cooling processes of the gas. We assume solar metallicity, corresponding to an elemental abundance of carbon and oxygen of \mbox{$1.4 \times 10^{-4}$} and \mbox{$3.16 \times 10^{-4}$}, respectively, relative to total hydrogen \citep{Sembach2000}. The chemical network is applied on-the-fly in the simulations, that is, the chemistry is evolved together with  the hydrodynamical evolution. In \citet{Ebagezio2023} we show that having an on-the-fly chemical network permits to obtain significantly more accurate chemical abundances with respect to post-processed chemical networks which assume chemical equilibrium, which is in agreement with findings of for example \citet{Glover2007b, Seifried2017, Hu2021, Seifried2022}.  The interstellar radiation field (ISRF) is set to \mbox{$G_0 = $ 1.7} in Habing units \citep{Habing1968} and the cosmic ray ionisation rate is \mbox{$3 \times 10^{-17}$ s$^{-1}$} with respect to atomic hydrogen. We also consider the attenuation of the ISRF using the \textsc{TreeRay/OpticalDepth} module \citep{Wunsch2018}. We consider simulations with and without magnetic fields. The magnetised runs have a magnetic field \textbf{\textit{B}} initialised to be parallel to the \textit{x}-direction with a strength of \mbox{$B_{x,0}$ = 3 $\mu$G} in the galactic mid-plane. The dust-to-gas ratio is fixed to be $1/100$ everywhere in the simulations. We refer to \citet{Walch2015} for further details concerning the initial conditions.
    
At the beginning of the SILCC simulations and up to a given time $t_0$ (see Table \ref{tab:overview_simulations}), supernova explosions drive turbulence. At $t_0$, further explosions are stopped and local over-densities are already visible. At this point, we focus on some of these over-dense regions (i.e. those which are part of the SILCC-Zoom project), zooming-in onto them with a resolution as high as 0.12~pc while keeping the larger-scale environment on a lower resolution. In this paper we use two HD and two MHD clouds. We refer to them as MC1-HD, MC2-HD, MC1-MHD, and MC2-MHD, respectively (see Table \ref{tab:overview_simulations}).

All runs include stellar feedback. We use sink particles to model the formation and evolution of stars \citep[see][for more details]{Haid2018,Haid2019}. The ionising radiation feedback from each massive star is treated with TreeRay \citep{Wunsch2021} and is fully coupled to the chemical network \citep{Haid2018,Haid2019}.
We note that currently we do not include non-ionising stellar radiation below 13.6~eV, which would add to the ISRF, that is $G_0$ (see also Appendix~\ref{sec:G0}).

\subsection{Chemical post-processing}
\label{sec:post_processing}

Before the actual synthetic emission maps are produced, we post-process the data to model physical phenomena which are not included in the simulation itself. In particular, we include: (i) CO freeze-out, (ii) splitting of H$_2$ into para- and ortho-H$_2$, (iii) microturbulence, (iv) \mbox{C$^+ \rightarrow $ C$^{2+}$} thermal ionisation, and (v) \mbox{C$^+ \rightarrow $ C$^{2+}$} photo-ionisation in regions affected by stellar feedback as a consequence of stellar radiation. The effects (i) -- (iv) are implemented in a pipeline\footnote{\url{https://astro.uni-koeln.de/walch-gassner/downloads/flashpp-pipeline}} developed by Pierre C. N\"urnberger, which couples the FLASH simulation output to RADMC-3D input required for the radiative transfer (Section~\ref{sec:radiative_transfer}).  As pointed out in \citet{Ebagezio2023}, the photo-ionisation (v) has a large impact on the abundance of C$^+$ within the HII regions and, as a consequence, also on the synthetic [CII] maps of those regions. It is thus essential to perform this post-processing step (v) in order to accurately model the [CII] within the HII regions. 
    
For this purpose, we sample the gas density, gas temperature, the stellar temperature and bolometric luminosity (the latter is obtained from the ionising radiation flux modelled in the 3D, MHD simulations) over the typical ranges occurring in our simulations, and build a grid of PDR models with \textsc{CLOUDY} \citep{Ferland2017}. For this post-processing step we use the entire frequency range also covering that of the ISRF, that is the local $G_0$ generated by nearby stars, to accurately calculate C$^{+}$ and C$^{2+}$ abundances. Next, using the information on the stellar content in our simulations, we obtain from this pre-calculated PDR database the updated fractional abundance of C$^+$ that takes into account the \mbox{C$^+ \rightarrow $ C$^{2+}$} photo-ionisation.  We then use it to replace in each cell the original abundance of C$^+$ given by \textsc{FLASH}. This update step is only done in regions where the 3D simulations predict the presence of ionising radiation coming from the stars. More details on this post-processing routine are provided in \citet{Ebagezio2023}.

\subsection{Radiative transfer}\label{sec:radiative_transfer}
    
We use the RADMC-3D code \citep{Dullemond2012} to do the radiative transfer calculations of the simulated clouds and to obtain synthetic emission maps. RADMC-3D is an open-source, 3D radiative transfer code, capable of performing both line and dust continuum radiative transfer calculations. We create synthetic emission maps of the [CII] 158 $\mu$m line and of FIR emission between 3 and 1100 $\mu$m. We define $t$ as the simulation time since the very beginning of the SILCC simulations and $\tevol$ as $t_\mathrm{evol} = t - t_0$. We consider the evolutionary stages at $\tevol = $ 2, 3, and 4~Myr for the HD simulations, and $\tevol = $ 3, 4, and 5~Myr for the MHD simulations, as the MHD clouds generally evolve somewhat slower \citep{Seifried2020, Ebagezio2023}. Table \ref{tab:overview_simulations} gives an overview of the simulations and the times at which the synthetic emission maps are produced. All emission maps are calculated for lines of sight (LOS) along the $x$-, $y$-, and $z$-axis. Furthermore, for MC1-HD and MC2-HD we produce maps at intervals of 100~kyr along the $z$-axis. We use these highly time-refined emission maps in Section \ref{sec:deficit_global_scale} to be able to better study the evolution of the considered characteristics.

    \begin{table}
        \caption{Overview of the simulations and snapshots used in this paper: run name, start time of the zoom-in, $t_0$, run type (HD or MHD), and the time $\tevol = t - t_0$ when synthetic emission maps are produced.}\label{tab:overview_simulations}
        \centering
        \begin{tabular}{cccc}
            \hline
            run name & $t_0$ [Myr] & run type &  $t_\mathrm{evol}$ [Myr] \\
             \hline
            MC1-HD & 11.9 & HD  & 2, 3, 4$^a$\\
            MC2-HD & 11.9 & HD  & 2, 3, 4$^a$ \\
            MC1-MHD & 16.0 & MHD  & 3, 4, 5 \\
            MC2-MHD & 16.0 & MHD  & 3, 4, 5 \\
            \hline
        \end{tabular}
        \tablefoot{$^a$For the LOS along the $z$-axis, snapshots in steps of 0.1~Myr are considered.}
    \end{table}
    
\subsubsection{Line radiative transfer}

In the following, we briefly describe the method to obtain the [CII] emission; for more details we refer to \citet{Franeck2018} and \citet{Ebagezio2023}.
In order to accurately model the [CII] line emission with RADMC-3D, we include microturbulence assuming that microturbulent broadening is as strong as the thermal broadening. Furthermore, we use the large velocity gradient (LVG) approximation \citep{Ossenkopf1997, Shetty2011a, Shetty2011b} to calculate the level population. We consider 201 equally spaced velocity channels for a velocity range of \mbox{$\pm 20$ km s$^{-1}$} around the [CII] rest frequency, which results in a spectral resolution of \mbox{d$v$ = 0.2 km s$^{-1}$}. This velocity range and spectral resolution enable us to capture the contribution of Doppler-shifted emission \citep{Franeck2018}. As we use the LVG approximation, we need to specify the collisional rates for C$^+$. For this, we rely on the Leiden Atomic and Molecular DAtabase \citep[LAMDA,][]{Schoier2005}. We consider para-H$_2$, ortho-H$_2$, H, and e$^-$ as collisional partners for C$^+$. 
We also model the effect of the Cosmic Microwave Background by assuming a background black body radiation at $T = 2.725$ K \citep{Fixsen2009},
which, however, only adds a negligible contribution both to the level population of C$^+$ and the obtained [CII] intensity.
 
\subsubsection{Dust continuum radiative transfer}

For calculating the dust continuum emission, the dust temperature $T_\mathrm{d}$ must be known. The \flash~simulation data contain information about dust temperature set by the ISRF and the collisional coupling of dust and gas. Its calculation does, however, currently not account for the radiative contribution of the stars formed in the simulations \citep[see][for our recent implementation to couple this radiation to the dust]{Klepitko2023}. Therefore, in a first step we use RADMC-3D to calculate the dust temperature resulting from the present stars using their luminosity and stellar surface temperature. In a second step, we consider for each simulation cell the maximum of the dust temperature given by \flash~and the one given by RADMC-3D. In this way, we properly account for the fact that dust can also be heated via energy transfer from the gas due to collisions as well as heating from the ISRF (which both are accounted for in FLASH). As this effect is not included in RADMC-3D, this could lead to rather low dust temperatures in regions with very low stellar radiation, whereas in regions with a strong stellar radiation, the dust temperature is mainly set by this radiation. Hence, by taking the maximum of the two available dust temperatures, we consider the effect of all dust heating sources in our $T_\mathrm{d}$ calculation.
        
RADMC-3D calculates $T_\mathrm{d}$ using a Monte Carlo method described in \citet{Bjorkman2001} with the continuous absorption method of \citet{Lucy1999}. For this purpose, information on absorption and scattering coefficients for different wavelengths are needed. These coefficients depend mainly on the composition and size of the grains. In general, the extinction coefficient $\kappa$ scales roughly as $\kappa \propto \lambda ^{-2}$ at long wavelengths, but varies significantly at shorter wavelengths. We use the values from \citet{Weingartner2001}, which aim to model a mixture of carbonaceous and silicate grains typical of MCs under solar neighbourhood conditions (their model B with $R_V$~=~4.0). For the calculation of $T_\mathrm{d}$ we use $10^{10}$ photon packages (see Appendix~\ref{sec:resstudy} for a resolution study).

Once the dust temperature is calculated, we use RADMC-3D to calculate the FIR dust continuum emission. We produce maps of 101 wavelengths between 3 and 1100 $\mu$m, equally spaced in logarithmic space. We then compute the integrated FIR luminosity by integrating in the range \mbox{$40 - 500$ $\mu$m} as this is used in different observational works \citep[e.g.][]{Sanders1996, Goicoechea2015, Pabst2021}, allowing for a straightforward comparison with observations.

\section{Results}\label{sec:results}

\begin{figure*}[ht]
    \centering
    \includegraphics[width=0.9\linewidth]{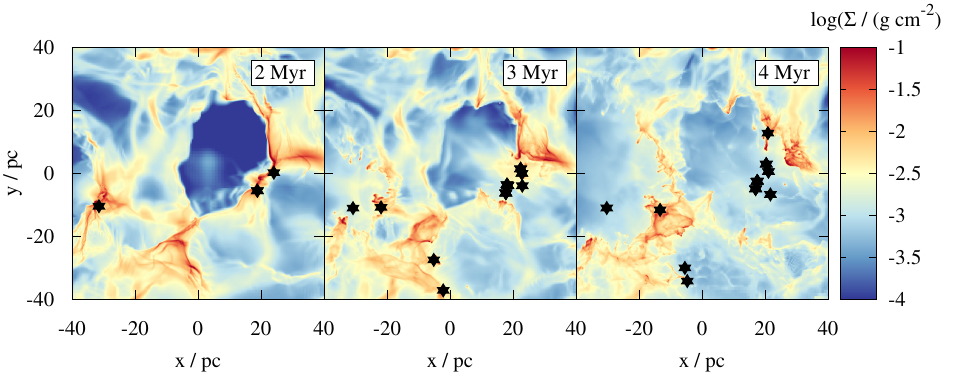}
    \caption{Column density maps of MC2-HD from $t_\mathrm{evol}$ = 2 to 4~Myr projected along the $y$-direction (from left to right) showing the effect of star formation (black stars) on the dispersal of the cloud.}
        \label{fig:column_density}
\end{figure*}

\begin{figure}
    \centering
    \includegraphics[width=0.9\linewidth]{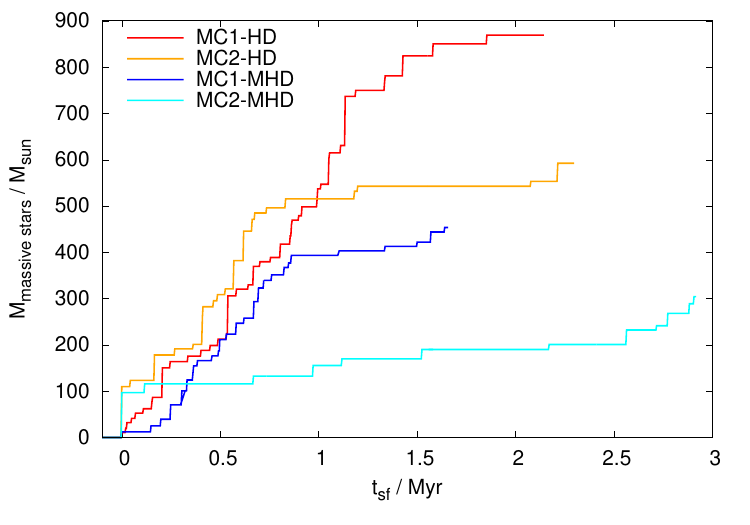}
    \caption{Time evolution of the mass of all massive stars ($M > 8$~M$_{\sun}$) formed in each MC. The clouds without magnetic fields form stars more rapidly.}
        \label{fig:star_formation}
\end{figure}

Before analysing the results concerning the [CII] and FIR emission, we give a brief overview of the general evolution of the clouds. For more details we also refer to \citet{Haid2019}. In Fig.~\ref{fig:column_density} we show the evolution of the column density for MC2-HD from $t_\mathrm{evol}$ = 2 to 4~Myr. The snapshots show the filamentary and complex gaseous structure of the MC. Over time, star formation sets in at various places in the cloud, which subsequently disperses the clouds by creating large HII regions which are almost completely devoid of any gas.

In Fig.~\ref{fig:star_formation} we show the mass of all massive stars ($M > 8$~M$_{\sun}$) formed for all four clouds considered in this work. In order to better compare different clouds, we here use the time $t_\mathrm{sf}$, which has passed since the first massive star has formed in a specific cloud at $t_\mathrm{sf,0}$, that is $t_\mathrm{sf} = t - t_\mathrm{sf,0}$ ($t_\mathrm{sf,0}$ is larger than $t_0$ used to define $t_\mathrm{evol}$, see Table~\ref{tab:overview_simulations}). Table \ref{tab:t_sf} shows $t_\mathrm{sf,0}$ and the relation between $t_\mathrm{evol}$ and $t_\mathrm{sf}$ for our clouds\footnote{We note that the first sink particle (including low mass stars) forms somewhat before $t_\textrm{sf, 0}$, as a massive star is only formed once the sink particle has reached a mass of 120~M$_{\sun}$ \citep{Haid2019}. However, as we only consider the feedback of massive stars  here, we use the formation time of the first massive star as a reference point.}. We note, however, that, unless explicitly specified, throughout the paper we use $t_\mathrm{evol}$.

For the clouds MC1-HD, MC2-HD and MC1-MHD there is a phase of rapid star formation within the first $\sim$1~Myr followed by a phase with a low star formation rate. Overall, for the clouds including magnetic fields, the star formation rate appears to be lower due to the stabilising effect of the magnetic field against gravity \citep{Girichidis2018, Seifried2020, Ebagezio2023,Ganguly2023}. Particularly, for MC2-MHD, star formation starts with a rather low rate, but continues over a longer time with an indication of acceleration towards late times.
    
    \begin{table}
        \caption{Time at which the first massive star ($M > 8$~M$_{\sun}$) forms (from the beginning of the simulation) and conversion between $\tevol$ and $t_\mathrm{sf}$.}
        \centering
        \begin{tabular}{ccc}
            \hline
            cloud & $t_\mathrm{sf,0}$ [Myr] & $t_\mathrm{evol} \rightarrow t_\mathrm{sf}$\\
             \hline
            MC1-HD & 13.75 & $t_\mathrm{sf} = \tevol - 1.85$ Myr \\
            MC2-HD & 13.60 & $t_\mathrm{sf} = \tevol - 1.70$ Myr \\
            MC1-MHD & 19.34 & $t_\mathrm{sf} = \tevol -3.34$ Myr \\
            MC2-MHD & 18.10 & $t_\mathrm{sf} = \tevol - 2.10$ Myr \\
            \hline
        \end{tabular}
    \tablefoot{For MC1-MHD, the first snapshot, at $\tevol = 3$ Myr, is before the onset of star formation.}
    \label{tab:t_sf}    
    \end{table}

\subsection{Dust temperature}
\label{sec:tdust}

    \begin{figure}
        \centering
        \includegraphics[width=0.99\columnwidth]{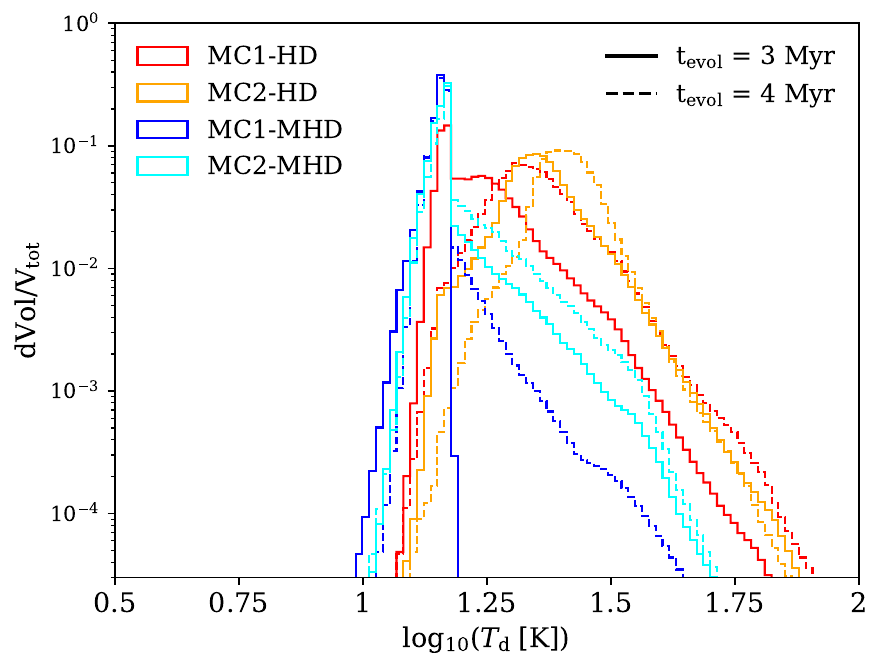}
        \caption{Dust temperature PDFs for all four clouds at $t_\mathrm{evol}$ = 3 (solid lines) and 4~Myr (dashed lines). There is a wide spread of temperatures over about 1 order of magnitude, which increases as star formation proceeds over time.}
        \label{fig:tdustpdf}
    \end{figure}
    
We next analyse the dust temperature, $\tdust$, of our MCs. This is important in order to understand the characteristics of the FIR continuum that we analyse in subsequent steps. First, we consider the distribution of dust temperatures of the four different clouds at $t_\mathrm{evol}$ = 3 and 4~Myr (Fig.~\ref{fig:tdustpdf}). Overall, there is a wide spread of observed dust temperatures from about 10~K to 100~K. Over time the spread increases and the dust becomes on average warmer due to the heating by newly formed stars. This effect is particularly pronounced for the MC1-MHD, where star formation is slightly delayed (see Table~\ref{tab:t_sf}).

    \begin{figure}
        \centering
        \includegraphics[width=0.99\columnwidth]{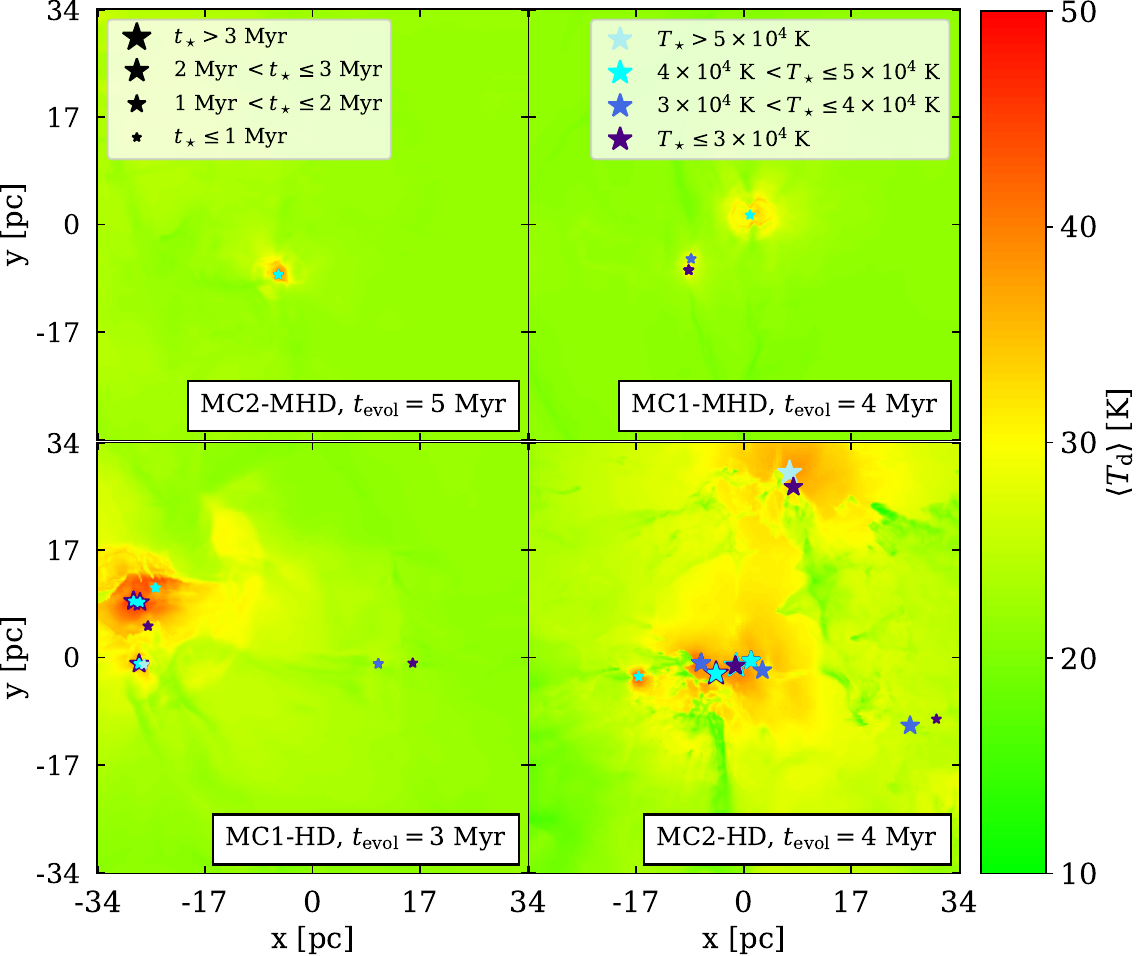}
        \caption{Maps of the mass-weighted mean dust temperature, $\langle \tdust \rangle$, for four snapshots of the simulated clouds. We show, from top left to bottom right, MCs with progressively more advanced star formation. We overplot the stars present in the simulations, where the size and colour indicate the age and temperature of the stars, respectively. In general, regions not affected by stellar feedback have \mbox{$\langle \tdust \rangle \, \simeq 15$ K}, and regions where stellar feedback is important have $\langle \tdust \rangle \, \simeq 50$~K. These regions are small (a few pc) and limited to the direct vicinity of the stars in the early stages of star formation (e.g.~MC2-MHD, top left), but involve larger parts of the cloud (a few 10 pc) at later stages (e.g.~MC2-HD, bottom right).}
        \label{fig:dust_temperature_map}
    \end{figure}

Next, to make predictions comparable to observations, in Fig.~\ref{fig:dust_temperature_map} we show four projected dust temperature maps, showing the mass-weighted mean dust temperature, $\langle \tdust \rangle$, averaged along the LOS. Contrary to Fig.~\ref{fig:tdustpdf}, we show snapshots for different values of $t_\mathrm{evol}$. In this way, the chosen four snapshots correspond to different stages of star formation in MCs. Moving from top left to bottom right, we show snapshots with progressively more advanced star formation and thus also more and more stars can be found. In general, we find \mbox{$\langle \tdust \rangle \simeq 15$ K} in regions not affected by stellar feedback, and \mbox{$ \langle\tdust \rangle \lesssim 50$ K} in regions where stellar feedback is relevant. Feedback-affected regions are confined to a small area of a few pc when star formation has set in recently (\mbox{$t_\mathrm{sf} \leq 1$ Myr}, top row of Fig.~\ref{fig:dust_temperature_map}), and are extended to a large portion of the cloud (a few 10~pc) when star formation has progressed over a few Myr and several massive stars have been formed (bottom row of Fig.~\ref{fig:dust_temperature_map}). 

We emphasise that the LOS averaging reduces the observed range of dust temperatures. The actual dust temperature in the vicinity of stars can be considerably larger than $ \langle T_\mathrm{d} \rangle$, reaching values as high as 100~K and above (see Fig.~\ref{fig:tdustpdf}). Overall, the reported dust temperatures match reasonably well with other numerical works. In projected maps, \citet{Ali2018} and \citet[][]{Inoguchi2024} find similar dust temperatures of a few 10~K in HII regions embedded into MCs, also reporting an expansion of regions of warm dust over a few~Myr as star formation proceeds. Furthermore, the higher values for $T_\mathrm{d}$ in the (unprojected) 3D distribution up to $\sim$100~K (see Fig.~\ref{fig:tdustpdf}) are also found in simulations by \citet{Kim2023} and analytical work of \citet{Sommovigo2020}, supporting the robustness of our results.

\subsection{[CII], FIR, and [CII]/FIR maps}
\label{sec:cii_deficit_overview}

    \begin{figure*}
        \centering
        \includegraphics[width=0.99\textwidth]{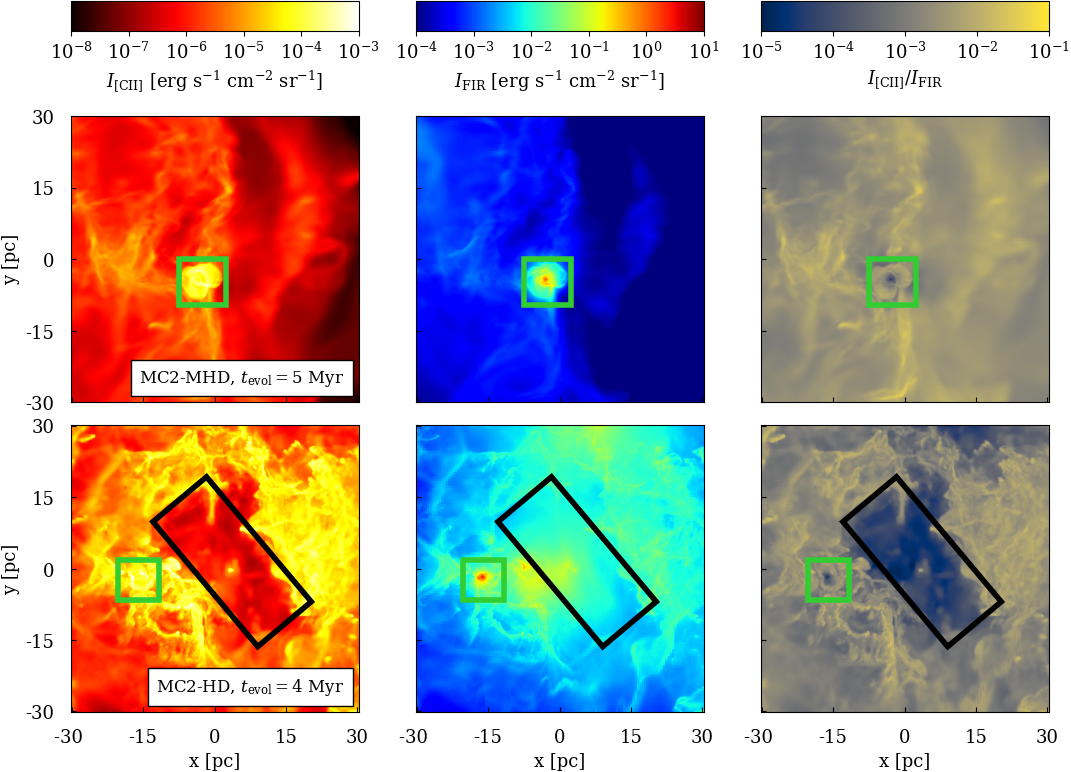}
        \caption{Emission maps of $I_\mathrm{[CII]}$ and $I_\mathrm{FIR}$, and maps of $\deficit$, for MC2-MHD, $\tevol = 5$ Myr (top), and MC2-HD, $\tevol = 4$ Myr (bottom). In the very young HII regions highlighted with a green rectangle, the low [CII]/FIR ratio is mostly due to the high $I_\mathrm{FIR}$. Conversely, the black rectangle highlights a more developed HII region, where the [CII] deficit is mainly caused by the \mbox{C$+ \rightarrow$ C$^{2+}$} ionisation due to stellar radiation.}
        \label{fig:cii_dust_deficit}
    \end{figure*}
    
Next, we analyse the [CII], FIR, and [CII]/FIR ratio maps. In short, we show that the [CII]/FIR ratio is particularly low in regions affected by recent massive star formation, and highlight two different mechanisms that can explain it.
    
To disentangle these two mechanisms, we show in Fig.~\ref{fig:cii_dust_deficit} maps of [CII], FIR and [CII]/FIR for two different snapshots. The left column shows the [CII] moment 0 maps, $I_\mathrm{[CII]}$, the central column gives the distribution of the FIR intensity, $I_\mathrm{FIR}$, and the right column shows $I_\mathrm{[CII]}/I_\mathrm{FIR}$. The top row corresponds to MC2-MHD at $\tevol = 5$~Myr and the bottom row corresponds to MC2-HD at $\tevol = 4$~Myr. These two snapshots were chosen as they depict the two typical reasons, which lead to a low [CII]/FIR ratio.

For discussing these two reasons, we focus on specific regions in these maps. First, we consider the region in MC2-MHD enclosed by the green square in the top row of Fig.~\ref{fig:cii_dust_deficit}. It contains a relatively young star (age of 0.33~Myr), which did not yet evacuate the gas in its surroundings, that is it did not yet create a low-density HII region. As a consequence, the heating due to its radiation causes a centrally concentrated peak of FIR emission with \mbox{$I_\mathrm{FIR} \simeq 5$ erg s$^{-1}$ cm$^{-2}$ sr$^{-1}$}. This is approximately three orders of magnitude above the values obtained for the surrounding cloud. Conversely, $I_\mathrm{[CII]}$ in that area is only larger by a factor of a few compared to the wider-spread,  high-intensity region extending to the left, as well as down- and upwards from the centre. As a result, $\deficit$ drops in the vicinity of the central star. 
    
A similar behaviour is found in the proximity of the stars on the left side of MC2-HD (bottom row, green square), where again $I_\mathrm{[CII]}$ is comparable to the emission coming from the surroundings, whereas $I_\mathrm{FIR}$ is significantly higher in the green box compared to its surroundings. Again, this leads to a lower $\deficit$ in that area than in its surroundings. We note that overall in MC2-HD $I_\mathrm{[CII]}$ is roughly an order of magnitude higher than in MC2-MHD. This is likely due to the stronger excitation of C$^+$ caused   by more free electrons acting as collisional partners and higher temperatures, which in turn are a consequence of the overall larger amount of massive stars in MC2-HD (see Fig.~\ref{fig:star_formation}).

    \begin{figure*}
        \centering
        \includegraphics[width=\textwidth]{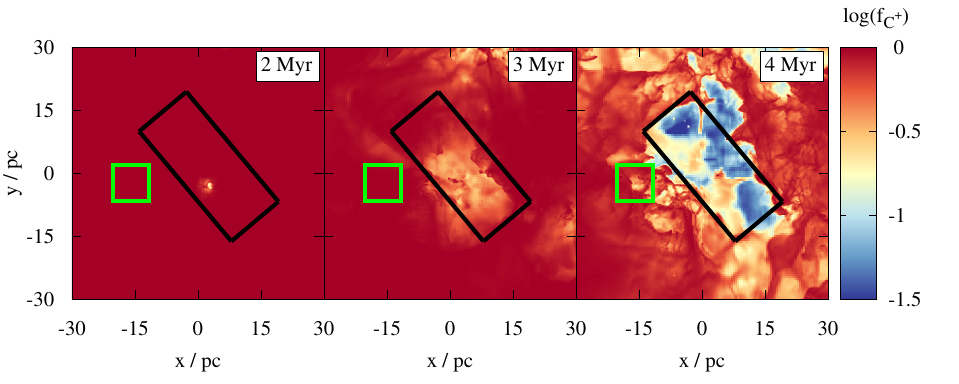}
        \caption{Fraction of C$^+$ (with respect to C$^+$ and C$^{2+}$) for MC2-HD at different times projected along the $y$-direction. As time evolves, C$^+$ gets progressively converted into C$^{2+}$ inside the HI regions due to further photo-ionisation by the radiation of the newly formed stars. This leads to a drop in the [CII] intensity, in particular in the region enclosed with the black rectangle (compare with Fig.~\ref{fig:cii_dust_deficit}).}
        \label{fig:fcplus}
    \end{figure*}

Beside the situation found in the two green squares in Fig.~\ref{fig:cii_dust_deficit}, there is another reason leading to a low [CII]/FIR ratio. This is indicated in the central part of MC2-HD (bottom row), enclosed by a black rectangle. The region corresponds to an evolved HII region, showing a large extent with low $\deficit$ values. Unlike for the other two regions, here this low ratio seems to be rather due to the lack of [CII] emission coming from the HII region, and less due to the enhancement of $I_\mathrm{FIR}$: there is only a moderately enhanced FIR emission, of the order of a few \mbox{$10^{-1}$ erg s$^{-1}$ cm$^{-2}$ sr$^{-1}$}, in the direct vicinity of stars, whereas the rest of the highlighted region is not distinguishable in FIR with respect to the other parts of the cloud. This is due to the dynamic effect of the ionising radiation coming from the more evolved, bright stars in the central part of the highlighted area. It sweeps up the surrounding gas, thus reducing the total column density in the HII region. Hence, even though $\langle \tdust \rangle$ is larger in that area (see Fig.~\ref{fig:dust_temperature_map}, bottom-right plot), $I_\mathrm{FIR}$ is relatively low. Furthermore, $I_\mathrm{[CII]}$ is very faint in this area (\mbox{$\simeq 10^{-6}$ erg s$^{-1}$ cm$^{-2}$ sr$^{-1}$}), comparable to the emission coming from very outskirts of the cloud. This is not only due to the aforementioned gas sweep-up, but mainly due to the second photo-ionisation $\mathrm{C}^+ \rightarrow \mathrm{C}^{2+}$ caused by the stellar radiation \citep[see Section~\ref{sec:post_processing} and][for details on the numerical implementation]{Ebagezio2023}.

This is demonstrated in Fig.~\ref{fig:fcplus}, where we show the (projected) fraction of C$^+$ with respect to the sum of C$^+$ and C$^{2+}$,
\begin{equation}
f_\mathrm{C^+} = \frac{N_\mathrm{C^+}}{N_\mathrm{C^+} + N_\mathrm{C^{2+}}} \, ,
\end{equation}
where $N_i$ is the column density of the corresponding carbon ion. As can be seen, at early stages there is almost no conversion of C$^+$ into C$^{2+}$, meaning \mbox{$f_\mathrm{C^+}$ $\simeq$ 1}. Contrary to that, at later stages a significant part of the C$^+$ has become further ionised. We find that due to the further photo-ionisation, the C$^+$ content is in parts reduced down to a few 1\% of the overall carbon content, in particular in the HII region marked with the black rectangle, where the [CII] intensity is particularly low (compare with Fig.~\ref{fig:cii_dust_deficit}). This clearly shows that a significant part of the reduction in $I_\mathrm{[CII]}$ is due to the further photo-ionisation of C$^+$ into C$^{2+}$.

We emphasise that the [CII] intensity in this region would be significantly higher without the further photo-ionisation of C$^+$. Not accounting for this would thus also affect the [CII]/FIR ratio, resulting in values which could be too high by up to a factor of $\sim$2.5 \citep[see figures~6 and A1 in][]{Ebagezio2023}.

Hence, we speculate that we identify a general evolution of the [CII]/FIR ratio as a function of the age of the associated HII region:
\begin{enumerate}
\item 
At early stages (which we marked with the green squares in Fig.~\ref{fig:cii_dust_deficit}), the reduction of the [CII]/FIR ratio occurs on small scales, driven by a significant enhancement of $I_\mathrm{FIR}$ in regions with newly formed stars.
\item 
At later stages (black rectangle in Fig.~\ref{fig:cii_dust_deficit}), and thus in more evolved HII regions, the reduction of the [CII]/FIR ratio involves larger scales and is caused mainly by the further photo-ionisation of C$^{+}$.
\end{enumerate} 
In order to investigate this more in detail, in Section ~\ref{sec:cii_deficit_local_scale} we systematically analyse  the $\deficit$ ratio  in all HII regions that we have in our simulated clouds.

\subsection{Projection effects}\label{sec:effects_of_line_depth}

    \begin{figure*}
        \centering
        \includegraphics[width=0.99\textwidth]{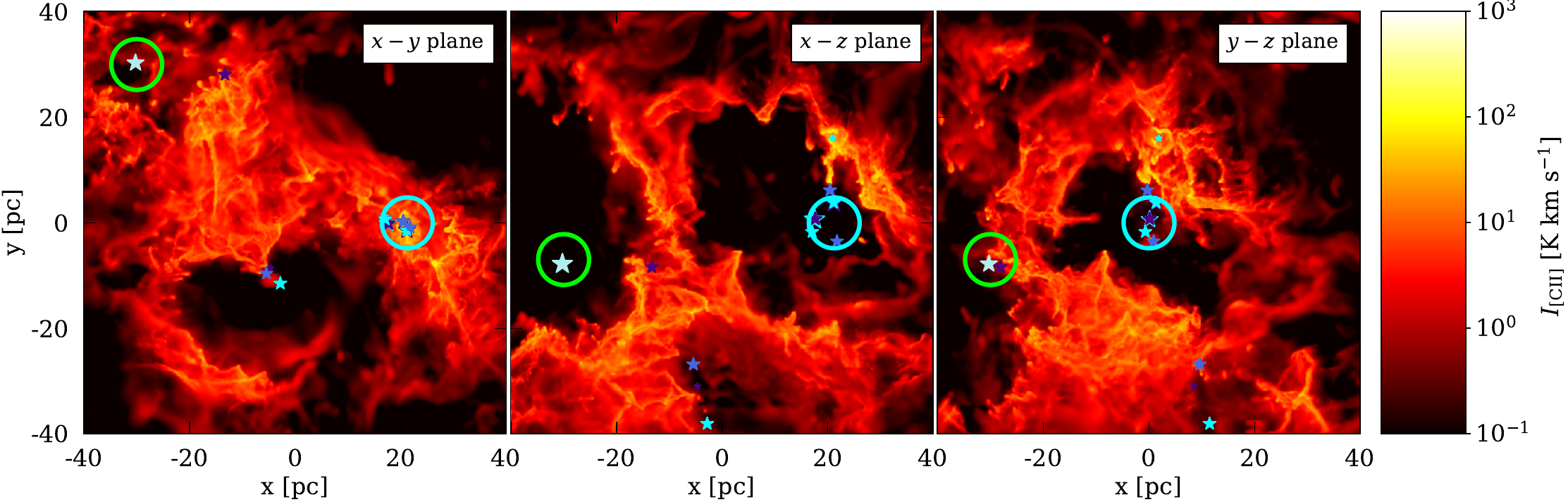}
        \caption{[CII] emission map of MC2-HD at $\tevol = 4 $ Myr for three different LOSs. The green and cyan circles highlight the stars (see Fig.~\ref{fig:dust_temperature_map} for a legend of the star symbols) responsible for the formation of `bubble 1' and `bubble 2', respectively. Because of the complex structure of the cloud, the shape of such HII regions is strongly dependent on the chosen LOS. As an example, ``bubble 2'' is not visible at all when looking along the $z$-axis (left panel). This explains the variations of the [CII] deficit as a function of the bubble radius for different LOS (Fig. \ref{fig:cii_deficit_all_bubbles}).}
\label{fig:bubbles_3_los}
    \end{figure*}

    \begin{figure}
        \centering
        \includegraphics[width=0.99\columnwidth]{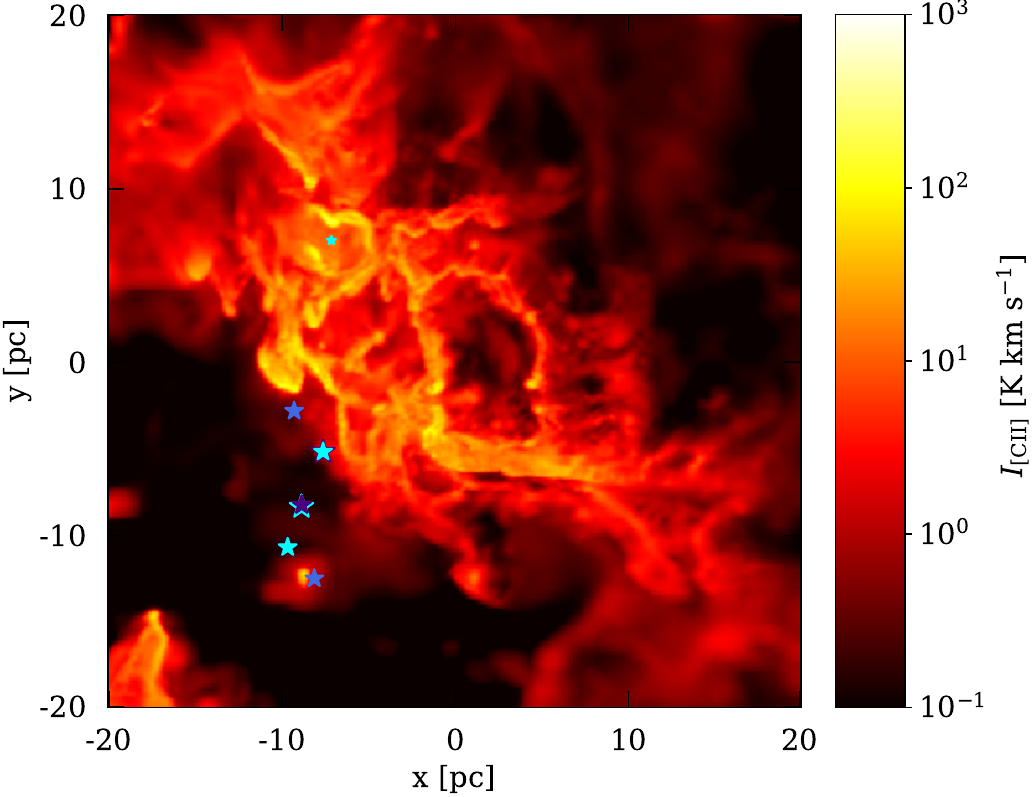}
        \caption{Zoom-in on an apparent cavity in the [CII] emission from the full [CII] emission map of MC2-HD at $\tevol = 4$~Myr shown in the right panel of Fig.~\ref{fig:bubbles_3_los}. This apparent cavity does not contain any stars (see Fig.~\ref{fig:dust_temperature_map} for a legend of the star symbols) and could be a potential consequence of projection effects, which can create apparent bubbles similar to expanding HII regions.}
        \label{fig:fake_bubble}
    \end{figure}

Before we further investigate the [CII]/FIR ratio, we first explore the importance of projections effects, which can be substantial due to the turbulent (sub) structure of MCs. In general, MCs are often observed to have filamentary \citep{Bally1987, Andre2014,Rezai2020,Hacar2023} and/or sheet-like substructures \citep{Zucker2021,Rezai2022}. This is verified for the simulations analysed in this paper, too \citep{Ganguly2023,Ganguly2024}. For this reason, the aspect ratio of a cloud is strongly dependent on the chosen LOS along which it is observed, which also affects the (apparent) shape of HII regions. For instance, a star-forming region in a sheet-like structure exhibits a ring shape when observed face-on, with the formed stars approximately in the centre. If the same HII region is observed edge-on, it may even be not observable at all due to the attenuation by the gas in the sheet (sitting in the foreground).

The impact of projection effects is evident in Fig.~\ref{fig:bubbles_3_los}, where we plot the $I_\mathrm{[CII]}$ emission maps of MC2-HD along the three principle axes. With green and light cyan circles, we highlight the stars responsible for the creation of two HII regions, ``bubble 1'' and ``bubble 2''. Indeed, the shape of the two bubbles is significantly different along the three LOS. In particular, bubble 2 cannot be seen at all in the \mbox{$x - y$} plane, but is clearly evident in the \mbox{$x - z$} and \mbox{$y - z$} planes. The bubble has a diameter of \mbox{$\sim 40$~pc} and is devoid of [CII] emission inside, due to the large amount of radiation coming from several hot stars inside (\mbox{$T_\mathrm{star} \lesssim 5 \times 10^4$ K}), which further photo-ionises C$^+$ to C$^{2+}$. Bubble 1 (associated with the stars in the green circle) exhibits different shapes when observed from different LOS as well, even though there is no LOS where the bubble is not visible at all. Nevertheless, the apparent size of the bubble changes significantly for different LOS.

Projection effects can also cause apparent bubbles, similar to those associated with expanding HII regions, but without any stars inside. Such an apparent bubble could be `created' by a dense foreground or background layer that separates the part of the real HII region containing the driving stars from a more distant region of the same region.  A possible example of such an apparent bubble is shown in Fig.~\ref{fig:fake_bubble}, a zoom-in of the right panel of Fig.~\ref{fig:bubbles_3_los}. In the centre of the figure there is a region with very low $I_\mathrm{[CII]}$ surrounded by a much brighter rim, which in combination resembles the structure of actual HII regions. However, there are no stars embedded in this structure. Hence, some caution is advised when analysing bubble structures in order to properly account for projections effects. We note that apparent bubbles with no stellar counterparts are also found in real observations of MCs and HII regions: for instance, \citet{Beuther2022} find several of such bubbles in [CII] emission maps of NGC 7538.

\subsection{The [CII]-FIR ratio within HII regions} \label{sec:cii_deficit_local_scale}
    
    \begin{figure*}
        \centering
        \includegraphics[width=0.99\textwidth]{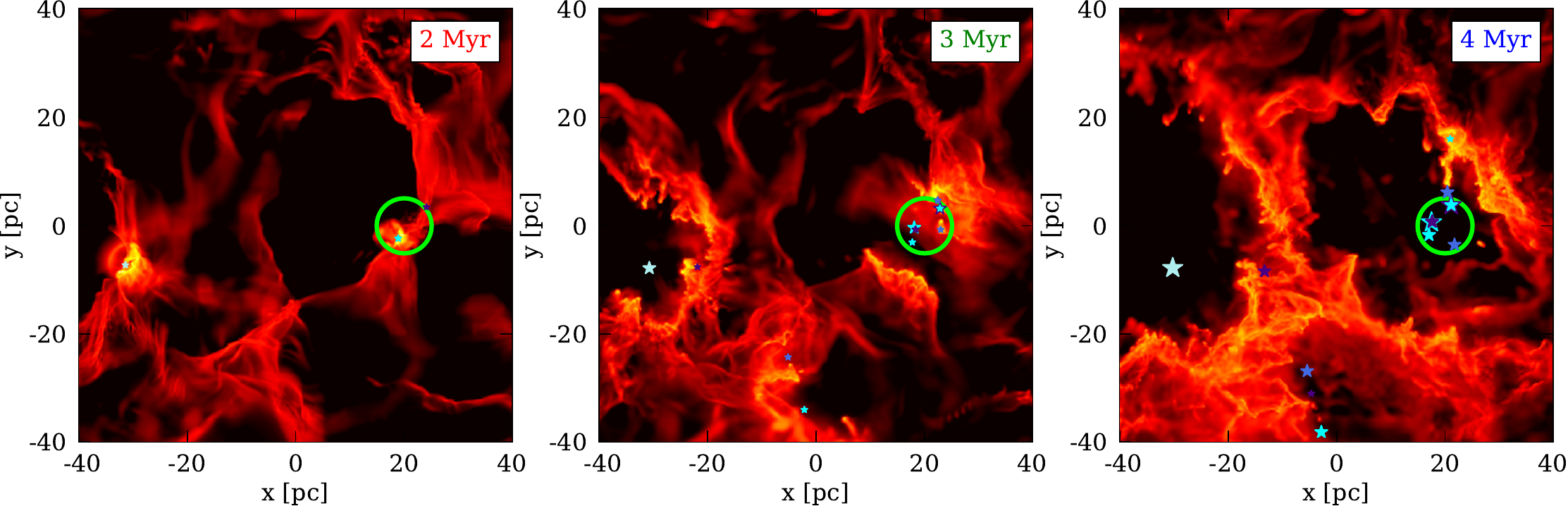}
        \includegraphics[width=0.99\textwidth]{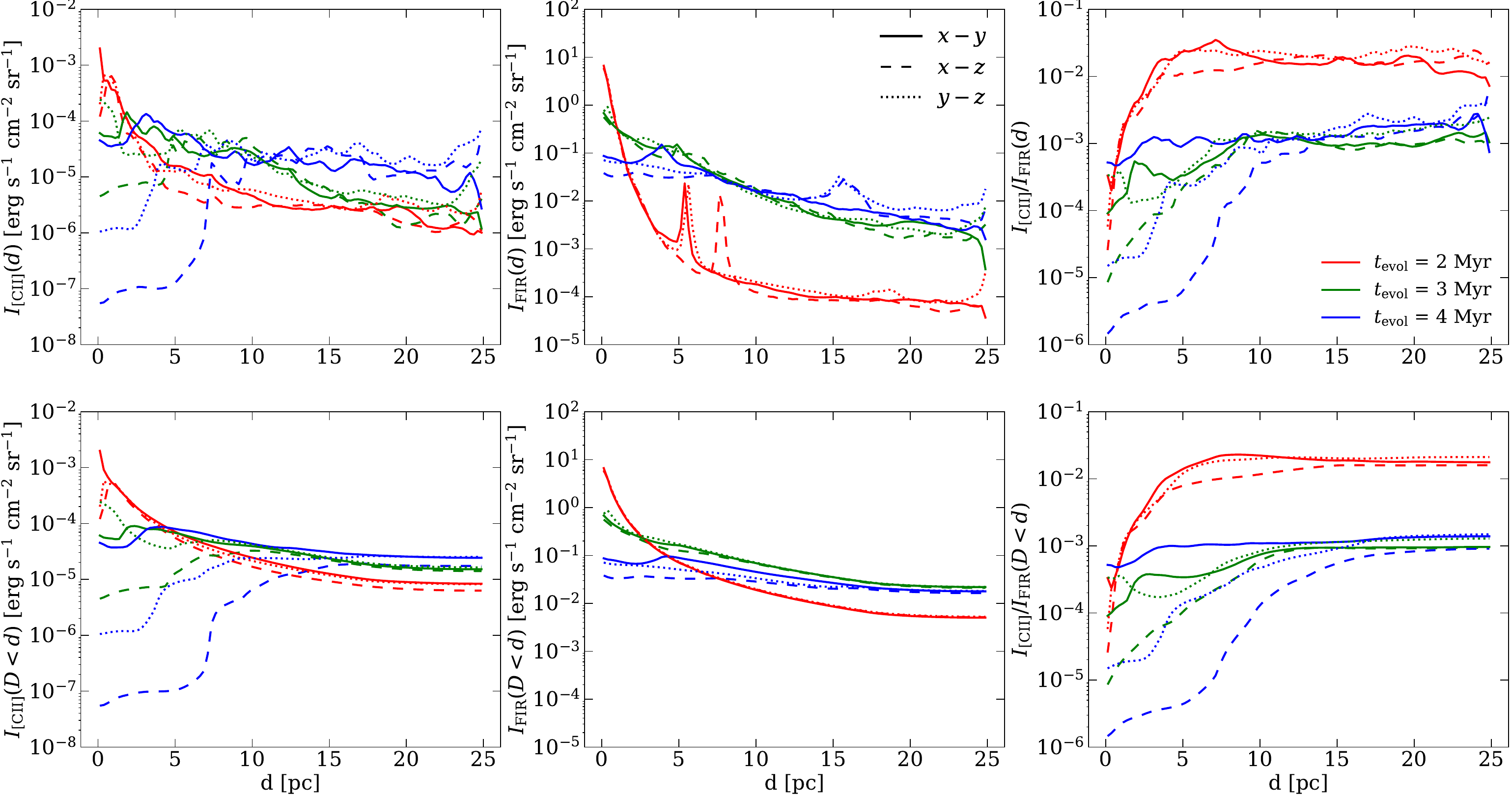}
        \caption{[CII] intensity maps and profiles for MC2-HD at different times. \textit{Top row}: Maps of $I_\mathrm{[CII]}$ for MC2-HD ($x-z$ plane) at $\tevol$ = 2, 3, and 4 Myr (from left to right). Stars are represented using the same symbols as in Fig.~\ref{fig:dust_temperature_map}. The green circles show the stars which are responsible for creating the HII region of interest, i.e. bubble 2 (see Table \ref{tab:main_star_in_snapshot}), inspected in the  middle and bottom row. \textit{Middle row:} average $I_\mathrm{[CII]}$, $I_\mathrm{FIR}$, and $\deficit$ at a given distance $d$ from the main star within the HII region highlighted by the green circles in the upper row. Solid, dashed, and dotted lines represent the values for the $x-y$, $x-z$, and $y-z$ plane, respectively. \textit{Bottom row:} same as in middle row, but now for an average over all points with a distance $D < d$.}
        \label{fig:cii_fir_deficit_vs_distance_mc2}
    \end{figure*}

    \begin{table}
        \caption{Cloud name and bubble name, evolutionary time, as well as the temperature and age of the main star (see main text for its definition) for all bubbles and snapshots analysed.}
        \label{tab:main_star_in_snapshot}
        \begin{tabular}{ccccc}
            \hline
            cloud & bubble & $t_\mathrm{evol}$ [Myr] & $T_\star$ [K]  & $t_\star$ [Myr] \\
             \hline
            MC1-HD & 1 & 2 & 27853 & 0.139 \\
             & & 3 & 46064 & 0.956 \\
             & & 4 & 45988 & 1.956 \\
            \hline
            MC1-HD & 2 & 3 & 25552 & 0.150 \\
             & & 4 & 42959 & 0.734 \\
            \hline 
            MC1-HD & 3 & 4 & 33551 & 0.310 \\
            \hline
            MC2-HD & 1 & 2 & 50137 & 0.317 \\
             & & 3 & 51357 & 1.317 \\
             & & 4 & 53643 & 2.317 \\
            \hline
            MC2-HD & 2 & 2 & 45983 & 0.152 \\
             & & 3 & 44513 & 1.152 \\
             & & 4 & 44202  & 2.152 \\
            \hline 
            {MC1-MHD} & {1} & {4} & {39970} & {0.276} \\
             & & {5} & {40934} & {0.965} \\
            \hline 
            MC1-MHD & 2 & 4 & 39498 & 0.330 \\
             & & 5 & 38795 & 1.330 \\ 
            \hline
            MC1-MHD & 3 & 4 & 40115 & 0.924 \\
             & & 5 & 39391 & 1.924 \\
            \hline
            MC2-MHD & 1 & 5 & 40085 & 0.33 \\
            \hline
        \end{tabular}
        \tablefoot{The main star is defined as the most luminous (therefore also the hottest) star within the oldest sink particle in a given bubble.}
    \end{table}
    
In this section we study $I_\mathrm{[CII]}$, $I_\mathrm{FIR}$, and $\deficit$ as a function of the distance from the centre of the HII region (see our definition below), in order to analyse radial trends of the [CII]-FIR ratio and the possible dependence on the age of the HII regions and on the kind of stellar population which creates them.
    
In Fig.~\ref{fig:cii_fir_deficit_vs_distance_mc2} we show an example of the procedure we follow. The top row shows, from left to right, the [CII] emission map of MC2-HD in the $x-z$ plane at $\tevol = $ 2, 3, and 4~Myr, respectively. By visual inspection of all snapshots for all times projected from the $x$-, $y$- and $z$-direction, we identify for each simulation various HII regions (or bubbles), which we list in Table~\ref{tab:main_star_in_snapshot}. We also identify their associated stars; for example in Fig.~\ref{fig:cii_fir_deficit_vs_distance_mc2}, we highlight with a green circle the stars which are responsible for the HII region of interest (namely bubble 2 of MC2-HD, see Table~\ref{tab:main_star_in_snapshot}).

Next, in the middle row of Fig.~\ref{fig:cii_fir_deficit_vs_distance_mc2} we show, from left to right, the azimuthal average of $I_\mathrm{[CII]}$, $I_\mathrm{FIR}$, and $\deficit$, taken in radial bins at a distance $d$, as a function of the distance from the centre of the bubble. We here define the centre of the bubble as the position of the ``main'' star of the HII region. It is defined as the hottest star in the oldest sink particle associated with the HII region (see Table~\ref{tab:main_star_in_snapshot}). This definition is motivated by the consideration that the size of the bubble is mainly determined by the time available for the expansion (this is the reason why we choose the oldest sink) and that the hottest star produces most of the radiation. We point out, however, that such a `main' star may not necessarily be the star which constitutes the main source of ionising radiation for all cells in the HII regions. This can happen when a bright star forms later on in a different sink particle. Hence, our definition of a main star is mainly used to define a geometric centre of the associated HII region for further investigation.

The bottom row shows the same quantities as the middle row, but for each distance $d$ we now average over all the pixels, which have a distance $D$ smaller than the given distance $d$:
    \begin{equation}
        I(D < d) = \frac{1}{N} \, \sum_{i = 1}^N I_i \; .
    \end{equation}
Solid, dashed, and dotted lines represent the calculation performed for the $x-y$, $x-z$, and $y-z$ plane, respectively. 
    
By looking at both, the emission maps and $I_\mathrm{[CII]}(d)$ (Fig.~\ref{fig:cii_fir_deficit_vs_distance_mc2}, top and middle row), we note that at an early evolutionary stage $I_\mathrm{[CII]}$ decreases with increasing $d$. When going to later evolutionary stages, $I_\mathrm{[CII]}$ is low at small $d$ (up to a few pc), then rapidly increases and eventually slightly decreases again at large $d$. This is a direct consequence of the expansion of the HII regions with time, as well as of the further ionisation of C$^+$ inside such regions (lowering $I_\mathrm{[CII]}$) and its high excitation in the rims (enhancing $I_\mathrm{[CII]}$). The distance within which such a lack of [CII] is observed, depends not only on the age of the HII region but also on the LOS considered (see e.g. the various blue lines for $\tevol = 4$ Myr), which can be attributed to the aforementioned projection effects (Section \ref{sec:effects_of_line_depth}).
    
The FIR intensity (central panel of the middle row) is very high at early evolutionary stages, reaching \mbox{$\sim$0.1~\intunity} in the proximity of the formed stars and then rapidly decreases with increasing $d$ due to the drop in both density and dust temperature as we go further away from the stars. At later stages, the dust density is less peaked around the star and the dust is also warmer at large distances due to the presence of other stars: this results in a less pronounced (but still existing) decreasing trend of $I_\mathrm{FIR}$ with increasing $d$. The additional peaks in $I_\mathrm{FIR}$ (those at \mbox{$d = 5 - 7$~pc}) are due to other neighbouring stars. 
    
The features of $I_\mathrm{[CII]}$ and $I_\mathrm{FIR}$ together lead to an $\deficit$ ratio which progressively increases with $d$ (right panel in the middle row of Fig.~\ref{fig:cii_fir_deficit_vs_distance_mc2}). The increase is sharp at early stages, with $\deficit$ reaching a roughy constant value of $\simeq 10^{-2}$ at \mbox{$d \simeq 5$~pc}. At later stages, there is a central region where $\deficit$ ranges from $\sim 10^{-6}$ to $\lesssim 10^{-3}$, depending on the snapshot and the LOS, followed by an increase up to values of $\simeq 10^{-3}$. This is due to the aforementioned lack of [CII] emission inside the HII region and due to the enhanced $I_\mathrm{FIR}$ emission in the part of the cloud surrounding it. 
    
The profiles of $I_\mathrm{[CII]}$, $I_\mathrm{FIR}$, and $\deficit$ averaged over areas with $D \leq d$ (bottom row of Fig.~\ref{fig:cii_fir_deficit_vs_distance_mc2}) follow approximately the same trend as those averaged in radial bins. This is due to the fact that, in the averaging process, the outer bins contribute with more points than the inner bins as they contain more pixels. However, high values of $I$ in the inner bins can still bring $I(D < d)$ to values above $I(d)$ at high $d$ (this is the case, for instance, for $I_\mathrm{FIR}$ at $t_\mathrm{evol} = 2$ Myr).
    
    \begin{figure*}
        \centering
        \includegraphics[width=0.99\textwidth]{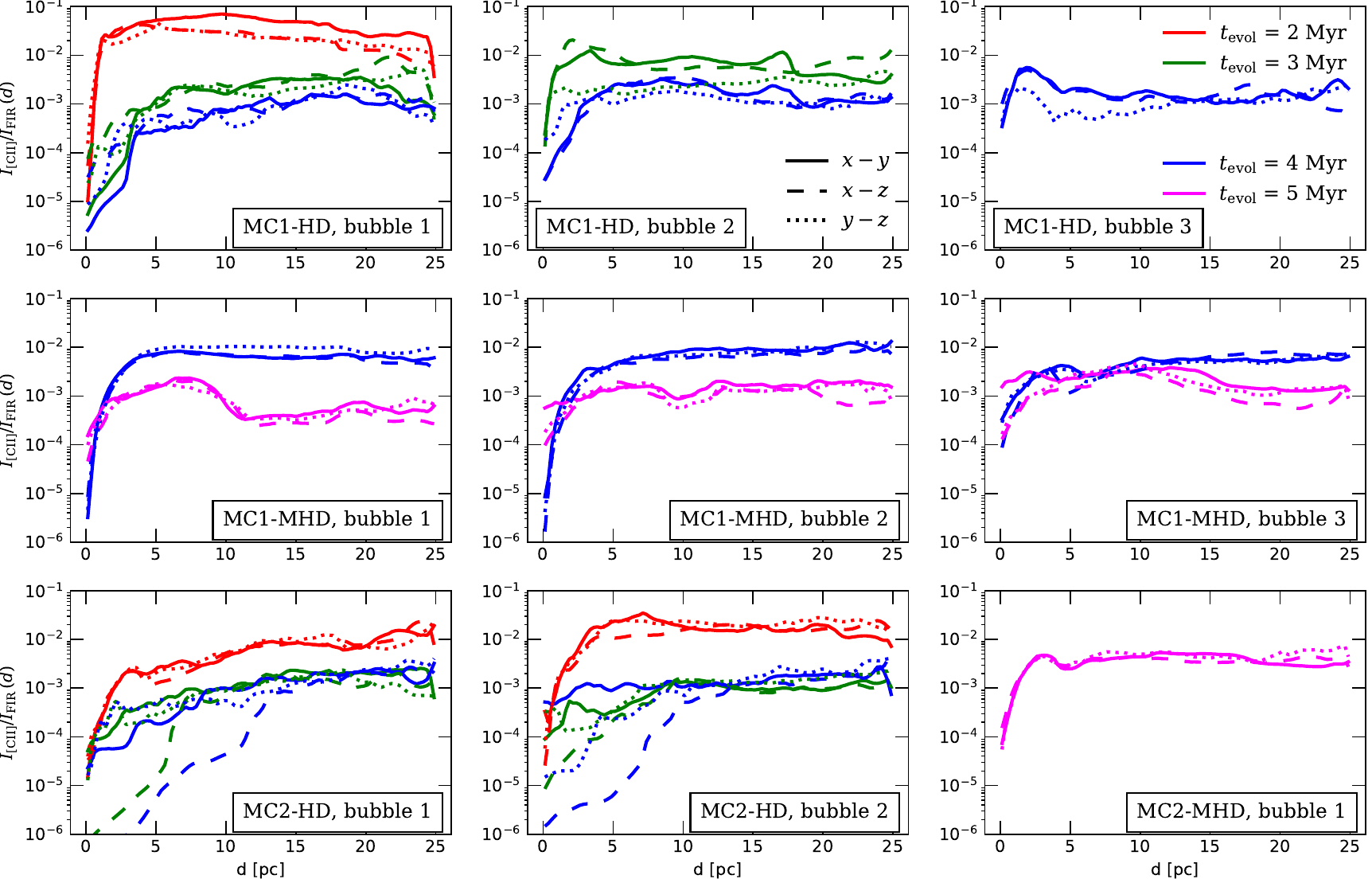}
        \caption{$\deficit$ as a function of the distance $d$ from the main star of all HII regions found in the investigated clouds. At small $d$, $\deficit$ drops to $10^{-6} - 10^{-4}$, depending on the regions and snapshots. At larger radii, the ratio is $\sim 10^{-2}$ at early stages, and $\sim 10^{-3}$ at later stages. The drop of the ratio at later stages is mostly due to the overall dust heating in the entire cloud caused by the various stars which have already formed.}
        \label{fig:cii_deficit_all_bubbles}
    \end{figure*}

The behaviour of $I_\mathrm{[CII]}$, $I_\mathrm{FIR}$, and $\deficit$ described above refers to a specific HII region within MC2-HD. However, the same physical phenomena, namely the dust heating, the clearing of the HII region due to ionising radiation, and the C$^+$ excitation and ionisation, produce similar trends in the other HII regions simulated here as well. However, differences in the stellar population and the surrounding environment can produce quantitative differences between those HII regions. In Fig.~\ref{fig:cii_deficit_all_bubbles} we show $\deficit$ as a function of $d$ for all HII regions identified in our simulations. For each bubble, we identify the main star of the HII region with the procedure described above.  We list the main star's properties for each bubble and time in Table~\ref{tab:main_star_in_snapshot}. We note that the stellar temperature, $T_\star$, evolves according to the model described in \citet{Gatto2017}, \citet{Peters2017} and \citet{Haid2019}. Comparing this table with the $\deficit$ plots in Fig.~\ref{fig:cii_deficit_all_bubbles}, we identify the following results:
    
\begin{itemize}
    \item At small distances (\mbox{$d < 2$ pc}), $\deficit$ drops to very small values as anticipated in Fig.~\ref{fig:cii_fir_deficit_vs_distance_mc2}. The lowest values reached are, however, independent of the main star temperature: for instance, for MC1-MHD at \mbox{$t_\mathrm{evol}$ = 4 Myr}, the main stars in the bubbles 2 and 3 have very similar temperatures, but the lowest values of $\deficit$ are $10^{-4}$ and $10^{-6}$, respectively.
    \item At intermediate distances (\mbox{$2 < d < 5 - 10$ pc}) the [CII]/FIR ratio quickly increases to values around $\sim 10^{-2}$ to $\sim 10^{-3}$. A [CII]/FIR ratio $< 10^{-3}$ extending up to radii of \mbox{5 - 10 pc} is only clearly visible in the bubbles 1 and 2 of MC2-HD and, to a smaller extent, in bubble 1 of MC1-HD. These three bubbles are characterised by an extremely hot star \mbox{($45\,000 - 50\,000$ K)} and generally a large number of stars inside them. This causes an extended HII region with a lack of [CII] emission, which is the reason of the low $\deficit$ ratio (Section~\ref{sec:cii_deficit_overview}). However, the geometry and the LOS from which we observe have a great impact on whether this feature is observable or not (see Section~\ref{sec:effects_of_line_depth}).  
    \item At large distances (\mbox{$d > 5 - 10$ pc}), $\deficit(d)$ is rather flat. Its value, however, decreases within approximately the first Myr of the HII region's lifetime from $\sim$10$^{-2}$ to $\sim$10$^{-3}$. This is a consequence not only of the ionising radiation of the main star, but also of the subsequent birth of numerous other stars in the HII region itself, as well as in other parts of the cloud (see top row of Fig.~\ref{fig:cii_fir_deficit_vs_distance_mc2}). The progress in star formation thus leads to an increased dust heating, increasing the FIR emission (central panel of Fig.~\ref{fig:cii_fir_deficit_vs_distance_mc2}) and thus lowering the $\deficit$ ratio.  
    \item Despite the qualitative trends of $\deficit$ with distance and time, we were not able to identify clear quantitative trends. 
    In general, we do not find a relation of the absolute value of $\deficit$ with neither the stellar age $t_\star$, its temperature or luminosity, nor the local column density evaluated within a radius of 2~pc (a proxy for the density in the surroundings of the star). We also did not find a relation of $\deficit$ with a combination of two out of the four aforementioned quantities. We speculate that the main reason for this is the fact that the $\deficit$ ratio is strongly influenced by a nonlinear interaction of all four quantities, projection effects as well as the exact density profile, which is not necessarily spherically symmetric.
\end{itemize}

    \begin{figure*}
        \centering
        \includegraphics[width=0.99\textwidth]{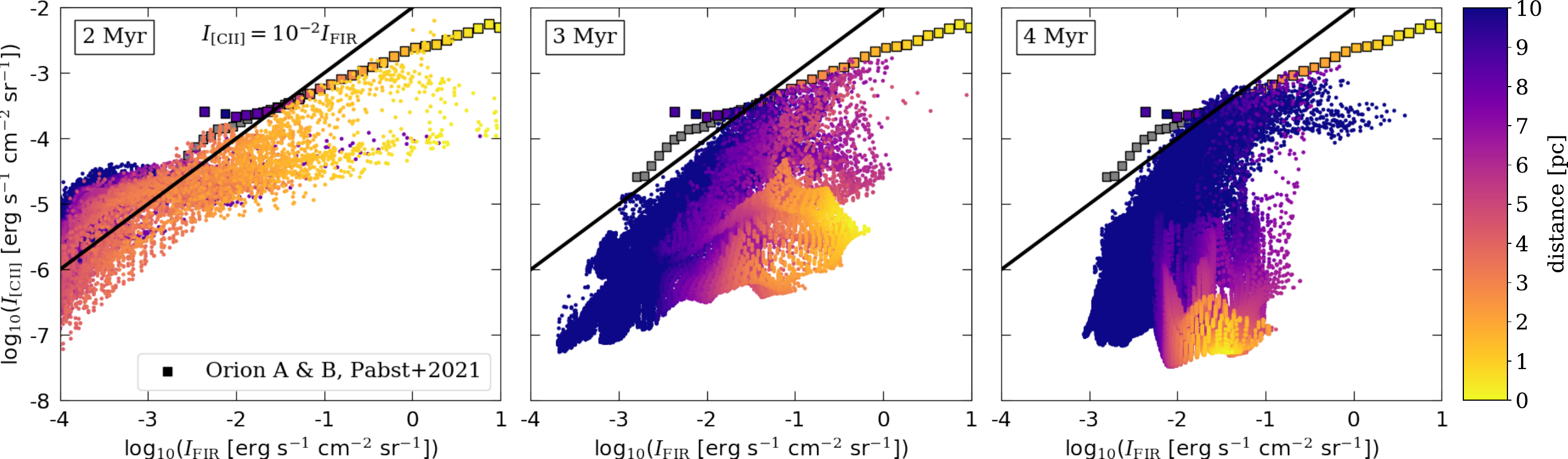}
        \caption{Distribution of $I_\mathrm{[CII]}$ as a function of $I_\mathrm{FIR}$ for all pixels within 10 pc from the main star of MC2-HD, bubble 2, in the \mbox{$x - z$} plane. The colour bar indicates the distance from the main star. To guide the readers eye, we show with the black line a constant ratio of \mbox{$I_\mathrm{[CII]}/I_\mathrm{FIR}$ = 10$^{-2}$}. For pixels at close-by distances, the values of $I_\mathrm{[CII]}$ and $I_\mathrm{FIR}$ decrease over time due to the evacuation of the bubble, whereas at larger distances the values remain in a similar range. This leads to a separation into a ``main branch'', formed by distant pixels, and a branch of close-by pixels, located below the main one. The squares show observational data from Orion A \& B taken from \citet[][binned for the sake of readability]{Pabst2021} for comparison.} 
        \label{fig:cii_vs_fir_colorbar_distance}
    \end{figure*}
    
Motivated by the analysis of Orion A presented by \citet{Pabst2021}, in Fig.~\ref{fig:cii_vs_fir_colorbar_distance} we show the distribution of $I_\mathrm{[CII]}$ and $I_\mathrm{FIR}$ for all pixels of one snapshot, colour-coded according to the distance to the main star. In all three panels we can identify a ``main branch'', characterised by a high $I_\mathrm{[CII]}$ for a given $I_\mathrm{FIR}$. At early stages \mbox{($\tevol = 2$ Myr)}, the pixels at small distance constitute a portion of this branch. However, as time evolves, the pixels at small distance are not on the ``main branch'' any more: they seem to form a separate branch which is located below the main one. This appearance of a second branch corresponds to the drop of the $\deficit$ ratio over time discussed before. The reasons for this are, as stated before, the stellar ionising radiation and the subsequent second ionisation of carbon. 

The ionising radiation disperses the cloud: therefore, the density of both dust and C$^+$ drop at small $d$. In combination with the second ionisation of C$^+$ into C$^{2+}$ (see Fig.~\ref{fig:fcplus}), this leads to the observed drop of $I_\mathrm{[CII]}$. For the FIR emission, the reduction of the density is partially compensated by the increase in dust temperature (see e.g. Fig.~\ref{fig:dust_temperature_map}), resulting in a smaller decrease in $I_\mathrm{FIR}$. Overall, this creates the separated branch, characterised by a lower $I_\mathrm{[CII]}$ for a given $I_\mathrm{FIR}$. For other bubbles and/or LOS we reproduce these two branches only partially. This is mostly due to projection effects (see also Section~\ref{sec:effects_of_line_depth}). 

In Fig.~\ref{fig:cii_vs_fir_colorbar_distance} we also overplot data for M42 in Orion~A and data for Orion~B obtained by \citet[][see their figure~5\footnote{Data kindly provided by C. Pabst (priv. comm.). For the sake of readability, we binned the data.
}]{Pabst2021}. The overall trend is similar showing an increase of  $I_\mathrm{[CII]}$ with $I_\mathrm{FIR}$ and larger values being found preferentially at smaller distances\footnote{We note that for Orion~A \citet{Pabst2021} only report datapoints up to distances of $\sim$5~pc, which is why do not expect observational datapoints at low values of $I_\mathrm{[CII]}$ and $I_\mathrm{FIR}$. For Orion~B they do not provide distance measures, which is why these datapoints are plotted with grey colour.}, although the observational data tend to lie slightly above our data points. 
This might be attributed to the potentially higher radiation field strength and density in this very compact region, which would lead to elevated intensities. For example, the luminosity of the binary system \mbox{$\theta^1$ Ori C} in Orion~A sums up to more than \mbox{$3 \times 10^5$~L$_\odot$} \citep{Gravity2018} compared to less than \mbox{$2 \times 10^5$~L$_\odot$} in the region simulated by us. In addition, projection effects and the generally highly complex density structures could cause deviations. Hence, given these potential differences, we  consider the agreement between the simulated and observed data as remarkable. In particular the slope for the high-intensity regions is similar, specifically for our early time at 2~Myr.

Finally, given that we see a slight decrease of $I_\mathrm{[CII]}$ at a given $I_\mathrm{FIR}$ over time in our simulations, this could mean that observationally found relations lying at lower $I_\mathrm{[CII]}$ belong to further evolved HII regions. However, given that (i) the spread in our simulated data is already rather large and (ii)  the aforementioned differences in the properties of the MC/HII region could potentially cause intrinsic variations, we consider the quantitative predictive power of such a time evolution as rather limited.

\subsection{The [CII] deficit on molecular clouds scales}
\label{sec:deficit_global_scale}

\subsubsection{Time evolution of the [CII] deficit}
\label{sec:CII_time_evolution}

    \begin{figure*}
        \centering
        \includegraphics[width=0.99\textwidth]{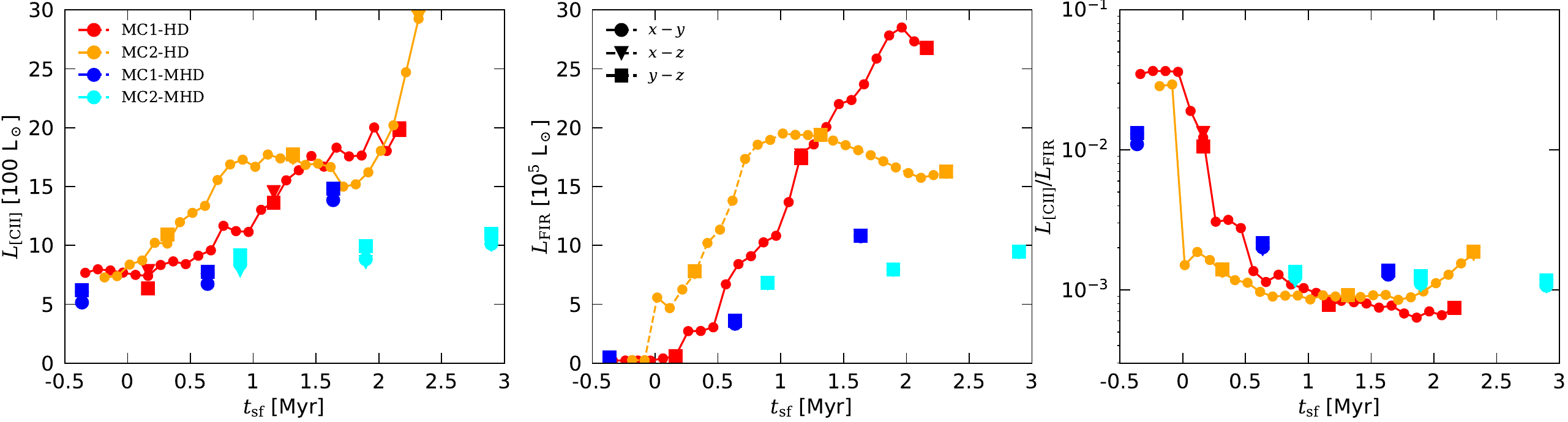}
        \caption{Evolution of $L_\mathrm{[CII]}$, $L_\mathrm{FIR}$, and $\deficitl$ (from left to right) as a function of $t_\mathrm{sf}$. The four clouds are represented with different colours and the three LOS with different symbols; the highly time-resolved datapoints along the $x-y$-projection for MC1-HD and MC2-HD are connected with a line. Overall, we observe a general tendency of a decreasing $\deficitl$ with increasing $t_\mathrm{sf}$. There is, however, some scatter in the values for different clouds at a given $t_\mathrm{sf}$, which we attribute to different star formation activities. We note the logarithmic $y$-axis scale in the right panel.}
        \label{fig:global_deficit_vs_tsf}
    \end{figure*}

Next, we investigate the [CII] and FIR luminosities (and their ratios) integrated over the entire cloud scale ($\lesssim$~100~pc) to mimic unresolved MCs in observations and study how they evolve over time. In Fig.~\ref{fig:global_deficit_vs_tsf} we show the evolution of $L_\mathrm{[CII]}$, $L_\mathrm{FIR}$, and $\deficitl$ with time, using $t_\mathrm{sf}$ as a measure, hence referring to the point when the first massive star forms (see Table~\ref{tab:t_sf}). We plot the results for all three LOS; note that differences among the different LOS are rather minor and hard to discern in the plot. For MC1-HD and MC2-HD, we add snapshots for the $z$-direction, which are separated by 0.1~Myr only, allowing us a more detailed time analysis.

The [CII] luminosity generally increases with time, as already pointed out in \citet{Ebagezio2023}. This is consistent with the common usage of $L_\mathrm{[CII]}$ as a star formation tracer \citep[e.g.][]{Herrera2015, Sutter2019, Bigiel2020, Pabst2021, Liang2024}. For MC1-HD, the increase in $L_\mathrm{[CII]}$ slows down after \mbox{$t_\mathrm{sf} \simeq 1.5$ Myr}. For MC2-HD, $L_\mathrm{[CII]}$ reaches a plateau at $t_\mathrm{sf} \simeq 1$ Myr, then slowly decreases up to $t_\mathrm{sf} \simeq 2$ Myr. This general behaviour matches well with the evolution of the mass of all massive stars in the simulations (compare with Fig.~\ref{fig:star_formation}). In addition, we see a second increase of $L_\mathrm{[CII]}$ for MC2-HD after \mbox{$t_\mathrm{sf} \simeq 2$ Myr}. We attribute it to the stellar radiation permeating now larger portions of the cloud due to its fractal structure including large, low-density areas \citep[compare figure~20 in][]{Seifried2020} and partially overlapping HII regions, leading to a larger amount of exited C$^+$. In addition, the stellar mass evolution in Fig.~\ref{fig:star_formation} also indicates a new star formation event in MC2-HD at late times.
For the two MHD clouds, the increase in $L_\mathrm{[CII]}$ is somewhat slower, which we attribute to the lower star formation rate in the beginning (Fig.~\ref{fig:star_formation}).

    \begin{figure}
        \centering
        \includegraphics[width=0.99\columnwidth]{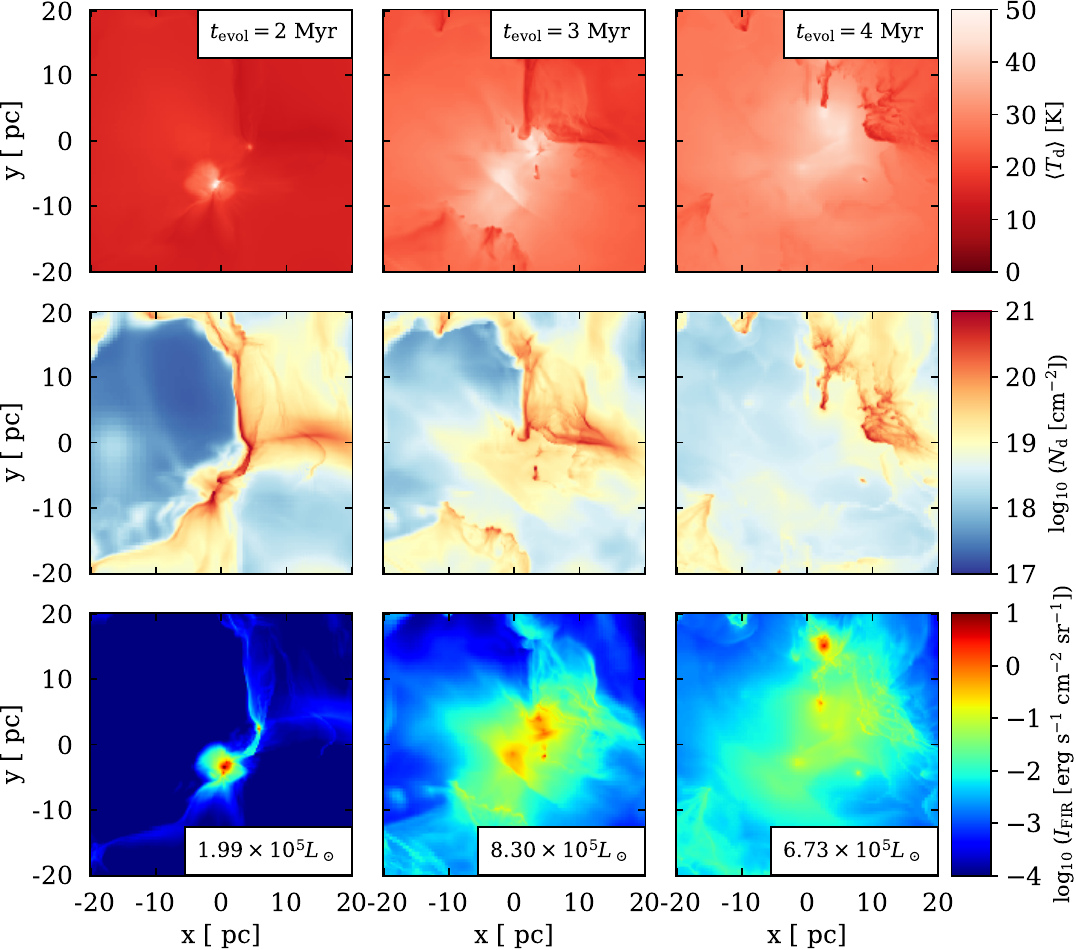}
        \caption{Mass-weighted dust temperature ($\langle \tdust \rangle$, top row), dust column density ($N_\mathrm{d}$, middle row), and FIR intensity ($I_\mathrm{FIR}$, bottom row) along the $y$-axis of a portion of MC2-HD centred around bubble 2 (see Table~\ref{tab:main_star_in_snapshot}) at $t_\mathrm{evol} = $ 2 , 3 and 4~Myr (from left to right). The expansion of the bubble progressively enlarges the region with warm dust, characterised by $\langle \tdust \rangle \gtrsim 20$~K. However, $N_\mathrm{d}$ in the warm region is reduced by the expansion of the bubble, which pushes the dust into the much cooler shell of the bubble. As a consequence, $I_\mathrm{FIR}$ arises from a progressively larger, but less FIR-bright area. This is also reflected in the total $L_\mathrm{FIR}$, reported for each snapshot in the $I_\mathrm{FIR}$ maps.}
        \label{fig:temperature_density_intensity_evolution}
    \end{figure}
    
Concerning the evolution of $L_\mathrm{FIR}$, we first note that for the two HD clouds, $L_\mathrm{FIR}$ increases rapidly, then reaches a maximum and finally starts decreasing again. The maximum occurs at \mbox{$t_\mathrm{sf} \simeq 2$}~Myr for MC1-HD and at \mbox{$t_\mathrm{sf} \simeq$1~Myr} for MC2-HD. The initial increase is caused by the bright FIR emission stemming from the area in the vicinity of the newly formed massive stars (indicated by the areas in the green rectangles in Fig.~\ref{fig:cii_dust_deficit}). The further evolution is indicated in the same figure, considering the bubble contained in the black rectangle. It shows a more evolved HII region, where there is more widespread FIR emission, but at a much lower intensity than at earlier stages.

We investigate the detailed evolution leading to this behaviour in Fig.~\ref{fig:temperature_density_intensity_evolution}, where we show the LOS-averaged, mass-weighted dust temperature $\langle T_\mathrm{d} \rangle$, the dust column density $N_\mathrm{d}$, and $I_\mathrm{FIR}$ for MC2-HD, bubble 2 (see Table~\ref{tab:main_star_in_snapshot}) at $t_\mathrm{evol} = $ 2, 3, and 4 Myr. Over time, there is a progressively larger region with contains relatively warm dust, namely with $\langle \tdust \rangle \gtrsim 20$~K. However, the warm region becomes less dense (middle row), as the stellar feedback pushes the dust towards the edge, forming a bubble. In the shell of the bubble, $\langle \tdust \rangle$ is measurably lower than inside.

Hence, there are two phenomena which counteract each other: (i) the enlargement of the warm regions leads to an increase of $L_\mathrm{FIR}$, and (ii) the redistribution of dust into the cooler shell leads to a decrease. 
The net result of these two processes is shown in the bottom row: there is a first phase when $L_\mathrm{FIR}$ increases, as the enlargement of the hot areas overcomes the decrease of $N_\mathrm{d}$ , and a second phase when $L_\mathrm{FIR}$ decreases again, as the redistribution (and cooling) of dust now takes over. 

This explanation so far refers to a single bubble. The evolution of $L_\mathrm{FIR}$ of an entire cloud can then be explained as the combined effect of several bubbles evolving in the manner described above. 
Specifically, for MC2-HD in Fig.~\ref{fig:global_deficit_vs_tsf}, we can see the phenomenon (i) followed by phenomenon (ii), which in turn is followed by phenomenon (i) attributed to a second phase of star formation (Fig.~\ref{fig:star_formation}).

Finally, we consider the $\deficitl$ ratio. Due to the stronger increase of $L_\mathrm{FIR}$ compared to that of $L_\mathrm{[CII]}$, the ratio decreases initially with $t_\mathrm{sf}$, down to a roughly constant value of $\sim$10$^{-3}$ with a scatter of a factor of a few. In MC2-HD, at $t_\mathrm{sf} \gtrsim 1.5$~Myr, $\deficitl$ increases again due to the aforementioned sharp increase of $L_\mathrm{[CII]}$. Overall, we can thus explain the origin of the [CII] deficit by the different qualitative evolution of the [CII] and FIR luminosity.
    
We note that the apparent slower increase of $L_\mathrm{[CII]}$ and $L_\mathrm{FIR}$ for the MHD clouds with respect to the HD clouds is due to the delayed evolution that the MHD clouds experience: magnetic fields slow down the star formation process as they act against gravitational collapse leading to a lower star formation rate \citep{Girichidis2018, Seifried2020, Ebagezio2023,Ganguly2023}.

\subsubsection{Dependence on H$^+$ and stellar luminosity}

   \begin{figure*}
       \centering
       \includegraphics[width=0.99\textwidth]{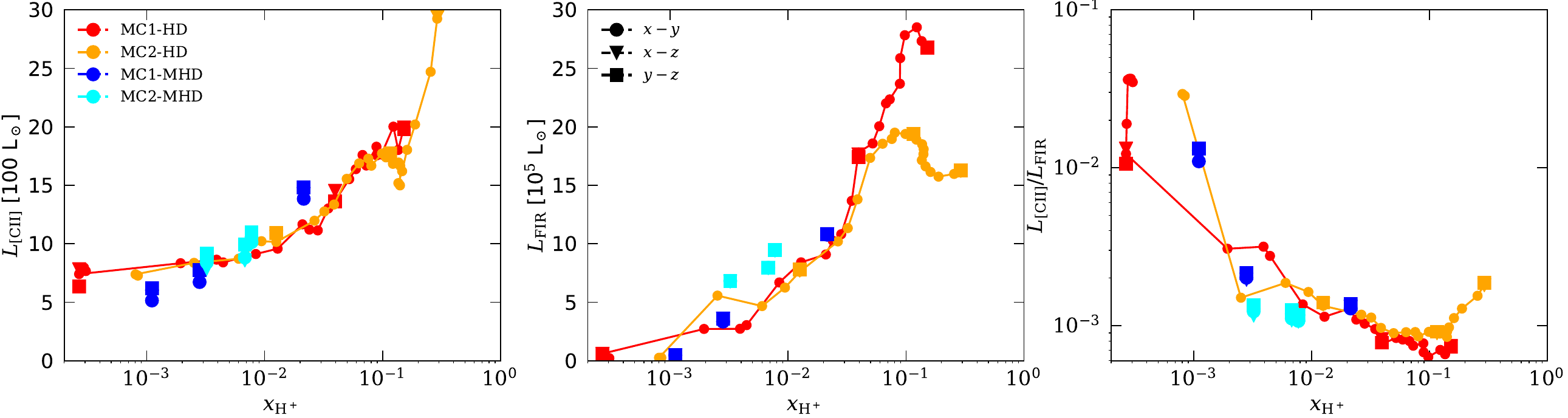}
       \caption{Same as Fig.~\ref{fig:global_deficit_vs_tsf}, but as a function of $x_\mathrm{H^+}$. We note a general increase of $L_\mathrm{[CII]}$ with increasing $x_\mathrm{H^+}$ and a maximum in $L_\mathrm{FIR}$ at \mbox{$x_\mathrm{H^+} \simeq 10^{-1}$} in the two HD clouds. As a consequence, $\deficitl$ decreases for increasing $x_\mathrm{H^+}$ up to this point, and then increases again.}
       \label{fig:global_deficit_vs_ionisation}
   \end{figure*}

   \begin{figure*}
       \centering
       \includegraphics[width=0.99\textwidth]{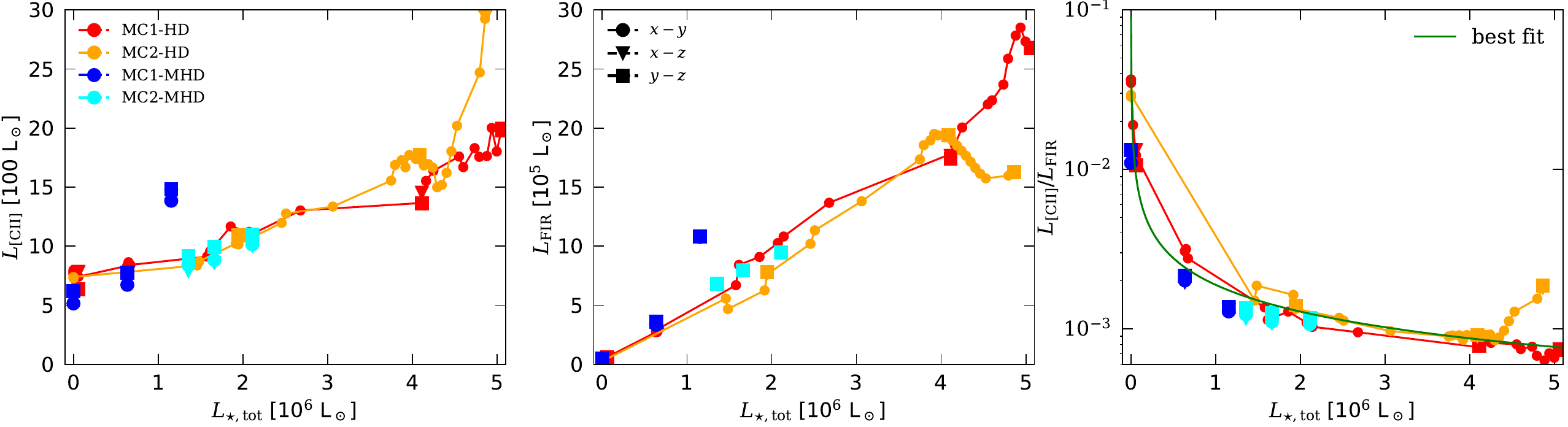}
       \caption{Same as Fig.~\ref{fig:global_deficit_vs_tsf}, but as a function of $\lstartot$. We fitted a power-law (Eq.~\ref{eq:bestfit}, green line) for the relation of $\deficitl$ with $\lstartot$. The relation holds as long as $L_\mathrm{FIR}$ increases (middle panel). After reaching its maximum, the decrease of $L_\mathrm{FIR}$, which has no counterpart in $L_\mathrm{[CII]}$, breaks this relation, which we therefore do not consider in the fit.}
       \label{fig:global_deficit_vs_star_luminosity}
   \end{figure*}   

Due to the creation of H$^+$ by stellar ionising radiation, we find a rough correlation between the mass of H$^+$ and $t_\mathrm{sf}$ in our simulations. For this reason, we next investigate the dependence on $\deficitl$ on the total mass fraction of ionised hydrogen. This is defined as 
\begin{equation}
    x_\mathrm{H^+} = \frac{M_\mathrm{H^+}}{M_\mathrm{H,tot}} \; ,
\end{equation}
where $M_\mathrm{H^+}$ is the total mass of H$^+$ in the cloud, and $M_\mathrm{H,tot}$ is the total hydrogen mass.

The dependence of $L_\mathrm{[CII]}$, $L_\mathrm{FIR}$, and $\deficitl$ on $x_\mathrm{H^+}$ is shown in Fig.~\ref{fig:global_deficit_vs_ionisation}. Symbols and colours have the same meaning as in Fig.~\ref{fig:global_deficit_vs_tsf}. Indeed, for each cloud the relations of $L_\mathrm{[CII]}$, $L_\mathrm{FIR}$, and $\deficitl$ with $x_\mathrm{H^+}$ are qualitatively similar to those with $t_\mathrm{sf}$. However, the differences among the clouds are smaller when $x_\mathrm{H^+}$ instead of $t_\mathrm{sf}$ is used. This suggests that $x_\mathrm{H^+}$ might be a better parameter to describe the evolution of $L_\mathrm{[CII]}$, $L_\mathrm{FIR}$ and $\deficitl$ than $t_\mathrm{sf}$. However, the correlation of $L_\mathrm{[CII]}$ with $M_\mathrm{H^+}$ is only of indirect nature: first, inside an HII region both the amount of C$^+$ (influencing $L_\mathrm{[CII]}$)  and H$^+$ are mainly set by a common factor, that is the amount of the stellar radiation. Second, in an HII region the excitation of C$^+$ is mainly due to free electrons, whose abundance in turn is directly proportional to the abundance of H$^+$, which is the main source of free electrons in HII regions. Also for $L_\mathrm{FIR}$ the coupling with $M_\mathrm{H^+}$ occurs only indirectly via the amount of stellar radiation, which both sets $M_\mathrm{H^+}$ and heats up the dust grains, thereby leading to an increase in $L_\mathrm{FIR}$.

In this context we note that H$_\alpha$ emission could further help to disentangle the apparent correlation between $M_\mathrm{H^+}$ and $L_\mathrm{[CII]}$. If $L_\mathrm{[CII]}$ is dominated by emission from within the HII region, this might lead to a good spatial correlation between the H$_\alpha$ and [CII] emission, whereas for the case that the [CII] emission comes mainly from the photodissociation region in the (neutral) shell, H$_\alpha$ and [CII] could be spatially disjunct. Overall, this might depend crucially on the evolutionary stage (Fig.~\ref{fig:cii_dust_deficit}) and/or the considered direction (Fig.~\ref{fig:bubbles_3_los}), which is why we defer a thorough investigation of this question to a subsequent work.

Next, we investigate the relation of $L_\mathrm{[CII]}$, $L_\mathrm{FIR}$, and $\deficitl$ with the stellar luminosity, $\lstartot$, as the impact of stellar feedback is determined, to some extent, by the stellar luminosity. We define $\lstartot$ as the sum of the bolometric luminosity of all stars with a mass \mbox{$M_\star > 8$~M$_\odot$}\footnote{As we do not explicitly model stars with masses lower than 8~M$_\odot$ in the simulations \citep{Haid2018,Haid2019}, for the moment we neglect those here. We argue, however, that due to the superlinear increase of luminosity with stellar mass, their contribution is rather minor.}. The relation of $L_\mathrm{[CII]}$ and $L_\mathrm{FIR}$ with $\lstartot$ (Fig.~\ref{fig:global_deficit_vs_star_luminosity}) is qualitatively similar to the relations with $t_\mathrm{sf}$ and $x_\mathrm{H^+}$, showing an initial increase and, in the case of MC2-HD, a subsequent slight decrease and a last sharp increase. However, the values for different clouds are more scattered, for a given $\lstartot$, than in the case of $x_\mathrm{H^+}$.

Similar results were also found by \citet{Vallini2017}, who modelled in a simplified manner the photoevaporation of MCs by stellar UV radiation and the resulting [CII] luminosity. Like here, the authors report a general increase of $L_\mathrm{[CII]}$ with $G_0$ (their parametrisation of the stellar luminosity). Also quantitatively our simulations and their models seem to reasonably match. Assuming a typical $G_0$ of $\sim$100 caused by the stellar radiation (see Fig.~\ref{fig:G0}), the authors obtain [CII] luminosities of a few 10$^3$~L$_\odot$ in good agreement with our findings. Furthermore, also the specific [CII] luminosities,  $L_\mathrm{[CII]}$/$M$, of 10$^{-(1.5 - 1.8)}$~L$_\odot/$M$_\odot$ reported by the authors for these $G_0$ values roughly agree with our findings of $L_\mathrm{[CII]}$/$M \simeq 10^{-(1.5-2)}$. Overall, the comparability of the findings obtained via these two different modelling approaches thus supports the robustness of the results reported here.
   
Lastly, the relation between $\deficitl$ and $\lstartot$ shows a decrease of $\deficitl$ for increasing $\lstartot$ for all clouds. For MC2-HD, it is followed by an increase at \mbox{$\lstartot \gtrsim 4 \times 10^6$ L$_\odot$}, which corresponds to the new star formation phase (see Fig.~\ref{fig:star_formation}) and the corresponding additional C$^+$ excitement already mentioned before. Remarkably, apart from the data points associated with this phase, the scatter of $\deficitl$ as a function of $\lstartot$ is minimal compared to the cases of $x_\mathrm{H^+}$ and $t_\mathrm{sf}$.  This relation can be well modelled with a power law, as shown in the right panel of Fig.~\ref{fig:global_deficit_vs_star_luminosity}. We fit a power law to all snapshots of MC1-HD and MC2-HD apart from those at very late stages, that is, those after $L_\mathrm{FIR}$ reaches its maximum ($t_\mathrm{sf} = 1$ Myr for MC2-HD and $t_\mathrm{sf} = 2$ Myr for MC1-HD). For this, we use the logarithmised values of  $L_\mathrm{[CII]}$ and  $\lstartot$ and fit a linear relation with the \textsc{polyfit} routine from \textsc{numpy} using equal weights for each data point and a $\chi^2$ minimisation. Translating this back into  a power-law relation (i.e. not-logarithmic data) yields the following best fit (green line in Fig.~\ref{fig:global_deficit_vs_star_luminosity}):
\begin{equation}
    \deficitl = 4.15 \times \left( \frac{ \lstartot}{L_\odot} \right)^{-0.56} \; .
\label{eq:bestfit}
\end{equation}
   
This fit provides a potential method to assess the total stellar luminosity of a cloud by using [CII] and FIR luminosity measurements. This relation holds up to when $L_\mathrm{FIR}$ increases. As we discussed in detail in Section \ref{sec:CII_time_evolution}, this corresponds to the phase of life of MCs, when stellar feedback heats the dust in the dense regions close to the star formation sites, but has not yet completely dispersed these regions and compressed the dust and gas into cool, dense shells surrounding the HII regions.

\section{Discussion}
\label{sec:discussion}

\subsection{Saturation of [CII] in HII regions}
\label{sec:cii_saturation}

    \begin{figure}
        \centering
        \includegraphics[width=0.99\columnwidth]{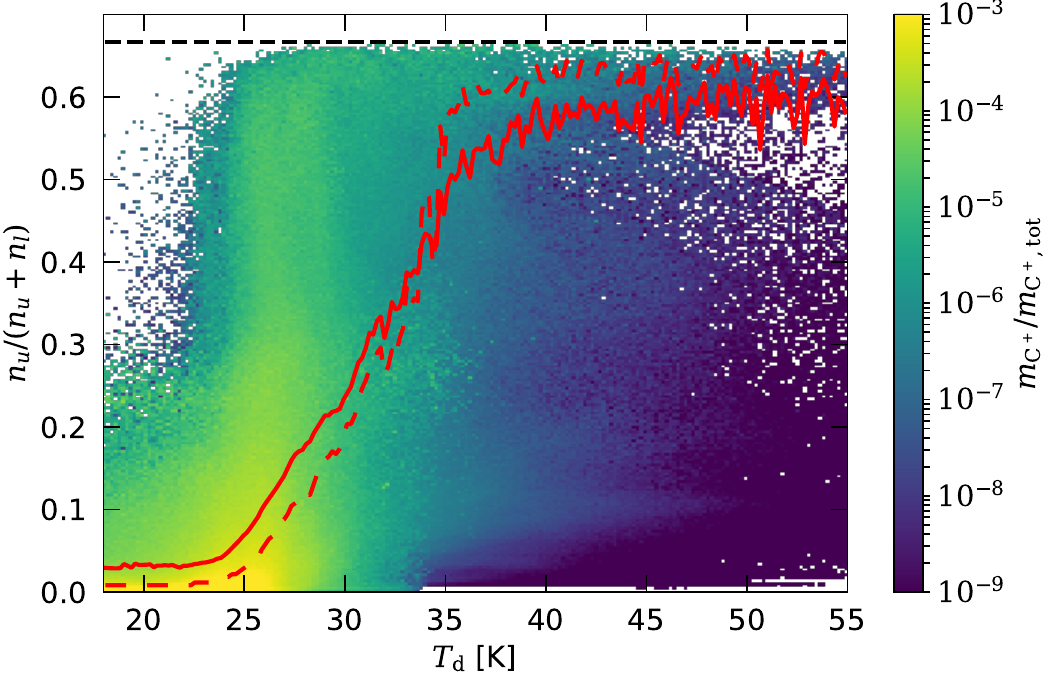}
        \caption{Mass-weighted 2D histogram of the fraction of C$^+$ ions in the $^2 P_{3/2}$ level ($n_u/(n_u + n_l))$ as a function of the dust temperature $T_\mathrm{d}$. The solid red line represents the mass-weighted average of $n_u/(n_u + n_l)$ for a given $T_\mathrm{d}$ bin, and the dashed line represents the median. The thin, dotted black line represents \mbox{$n_u/(n_u + n_l) = 2/3$}, that is, the maximum possible value. We show the result for MC2-HD at \mbox{$t_\mathrm{evol} = 4$ Myr}, with \mbox{$18 \; \mathrm{K}\leq T_\mathrm{d} \leq 55 \; \mathrm{K}$}. On average, C$^+$ reaches rough saturation at \mbox{$T_\mathrm{d} \simeq 40$ K}. This confirms that, in vicinity of stars, the [CII] line is saturated, causing, in part, the [CII] deficit.}
        \label{fig:lev_population}
    \end{figure}

The origin of the [CII] deficit has been often attributed to the saturation of [CII] caused by stellar radiation, which prevents the [CII] emissivity from increasing above a certain amount, no matter how strong the radiation field becomes \citep[][]{Kaufman1999, Munoz2016, Rybak2019, Bisbas2022}. Also \citet{Vallini2017} report in their idealised models of photoevaporating MCs a saturation of $L_\mathrm{[CII]}$ for radiation field strengths corresponding to $G_0 \gtrsim$ 10$^3$.
    
For the purpose of testing this hypothesis, we use RADMC-3D to calculate the level population of the C$^+$ ions (after correction for possible ionisation into C$^{2+}$) and analyse its dependence on the dust temperature, in order to link it to the FIR emission. More specifically, let $n_u$ be the number density of C$^+$ ions in the upper level, $^2 P_{3/2}$, and $n_l$ the number density of those in the lower level, $^2 P_{1/2}$. We then investigate the fraction of ions in the upper level, $n_u/(n_u+n_l)$ as a function of $T_\mathrm{d}$. In Fig.~\ref{fig:lev_population} we show this relation for MC2-HD at \mbox{$t_\mathrm{evol} =  4$ Myr} in form of a mass-weighted phase diagram. We overplot the mean and the median (solid and dashed red line, respectively). We do not consider cells with $T_\mathrm{d} < 18$~K, as here the main heating source is the ISRF, in which we are not interested in here. Furthermore, as for $T_\mathrm{d} > 55$~K, the number of cells is very low, which makes the mean and median very noisy, we also do not consider them here.
    
The fraction of ions in the excited state remains low \mbox{($n_u/(n_u+n_l) \simeq 0.03$)} up to $T_\mathrm{d} \simeq 23$ K, then slowly increases until \mbox{$T_\mathrm{d} \simeq 40$ K}, where the distribution appears to flatten around $\sim$0.6. We point out that the theoretical maximum of $\left(n_u / n_l \right)$ is 2, because of the statistical weights of the upper and lower level, and therefore \mbox{max$(n_u/(n_u + n_l)) = 2/3$}. This value is only reached for very high gas temperatures and when the density is above the critical density for C$^+$ of around 10$^{3}$~cm$^{-3}$. Hence, we do not expect $(n_u/(n_u + n_l))$ to exactly reach the maximum of 2/3 in our simulations. Nonetheless, the clear sign of flattening of $(n_u/(n_u + n_l))$ for $T_\mathrm{d} \geq 40$~K strongly indicates that the [CII] line indeed saturates in this regime. Since the regions with high $T_\mathrm{d}$ are those close to bright stars (Figs.~\ref{fig:dust_temperature_map} and \ref{fig:temperature_density_intensity_evolution}, top row), this shows that in the vicinity of bright stars the [CII] line is indeed saturated. 

Hence, overall the saturation of the [CII] line contributes to the rather low $\deficitl$ ratio, as most of the FIR emission comes from the vicinity of the stars -- where [CII] is saturated. In addition, as stated before, also the conversion of C$^+$ into C$^{2+}$ contributes to the $\deficitl$ low ratio, lowering $L_\mathrm{[CII]}$ by up to a factor of $\sim$2.5 compared to the case when no further ionisation would be considered \citep[][]{Ebagezio2023}.

\subsection{The [CII] deficit and its link to evolutionary stages of star formation}
\label{sec:CIIdeficit_evol}

In the following we summarise our results discussed so far concerning the origin and evolution of the [CII] deficit. Particularly, we link it to the different stages of MCs and star formation. We suggest the following correlation between star formation and the [CII] deficit:
\begin{enumerate}
    \item \textit{Phase~1: Pristine MCs:}
    Before star formation sets in, we see widespread and diffuse [CII] and FIR emission with typical intensity ratios of $\gtrsim 10^{-2}$ (regions outside the coloured boxes in Fig.~\ref{fig:cii_dust_deficit}). This results in an overall $\deficitl$ ratio averaged over the entire clouds of $\gtrsim 10^{-2}$.
  
    \item \textit{Phase~2: Onset of star formation and young HII regions:} Directly after new stars were formed,  the dust temperature and density in their surroundings are large, causing a high $I_\mathrm{FIR}$ in these regions. Simultaneously, the increase in $I_\mathrm{[CII]}$ is limited in this phase, due to the saturation of [CII] and the second ionisation of C$^+$ into C$^{2+}$. In combination, this causes the $\deficit$ ratio to be low close to the newly formed stars (see green enclosed regions in Fig.~\ref{fig:cii_dust_deficit}), whereas in the surrounding regions the $\deficit$ ratio still remains around 10$^{-2}$. Over time, the regions affected by the radiative feedback expand leading to a global decline of the $\deficitl$ ratio in MCs to values around $\sim10^{-3}$.

    \item \textit{Phase 3: Evolved HII regions:} In this phase, $L_\mathrm{FIR}$ remains constant or even decreases as the dust is pushed into cold shells while the HII regions are evacuated (region encapsulated with the black rectangle in Fig.~\ref{fig:cii_dust_deficit}).
    The elevated excitation of C$^+$ by free electrons and higher temperatures created by the stellar radiation, which now permeates the entire cloud, compensates the decrease of $I_\mathrm{[CII]}$ inside the evacuated HII region. Overall, this leads to a phase with roughly constant values of $\deficitl$ around $\sim 10^{-3}$.

    \item \textit{Phase~4: Secondary star formation:}
    The potential onset of a secondary star formation phase can lead to a subsequent increase of both $L_\mathrm{[FIR]}$ and $L_\mathrm{[CII]}$, where the latter appears to be more pronounced. This results in a slight increase of $\deficitl$ as seen for MC2-HD (see orange dots in Fig.~\ref{fig:global_deficit_vs_star_luminosity}).
\end{enumerate}

We emphasise that throughout phase~2 to 4, the $\deficitl$ ratio is lower than in a cloud not affected by stellar radiation (phase~1). From  Fig.~\ref{fig:global_deficit_vs_tsf} it can be seen that this stage is reached within short timescales of $\lesssim$1~Myr, which is typically significantly shorter than the duration of the star forming phase of clouds, during which they get disrupted. As shown in \citet{Haid2019}, for the simulations underlying this publication, the ionising feedback disrupts the clouds within a few Myr \citep[see also][]{Walch2012}. These timescales for cloud disruption are in line with a number of other numerical as well observational results, which are discussed in detail in the review of \citet[][see their chapters 4.2 and~5 for a summary and an extensive list of literature]{Chevance2023}. Hence, during most of their star forming phase, MCs will exhibit a rather low value of $\deficit$.

Hence, the (relatively seen) long timespan, at which star forming clouds exhibit a low $\deficit$ ratio around $\sim$10$^{-3}$, could thus explain the observed [CII] deficit \citep{Malhotra1997, Luhman1998, Malhotra2001, Luhman2003, Stacey2010,Casey2014, Smith2017, Hu2019}.

\subsection{Comparison with [CII] deficit observations}

We next compare our results with actual observations. Starting on scales of (and below) individual clouds, a well-studied MC in the solar neighbourhood is the Orion Molecular Cloud~1 (OMC~1). From the Sun, it is located beyond the Orion Nebula cluster at a distance of $\simeq 414$ pc \citep{Genzel1989}. It is an actively star forming cloud, similar to the ones we simulate here, despite being more massive. A study of the [CII] deficit in OMC~1 has been performed by \citet{Goicoechea2015}, which used Herschel HIFI data to obtain a $\deficit$ map. The authors measure the radial change of $\deficit$ with distance to their reference centre (the HII region around the Trapezium cluster). On the smallest scale of 0.1~pc, the authors obtain a value of \mbox{$9.3 \times 10^{-4}$}, which increases to \mbox{$3.8 \times 10^{-3}$} when averaged over the entire cloud (see their table~2). Similar values are found by \citet{Jackson2020} in their analysis of 4 star-forming clumps in the Milky Way (scales  of $\sim$1~pc), reporting on average \mbox{$\deficitl \simeq 10^{-4} - 10^{-3}$}. Generally, these observational values agree well with our results shown in Fig.~\ref{fig:cii_deficit_all_bubbles}, despite a certain underlying scatter. As discussed in Section~\ref{sec:cii_deficit_local_scale}, this scatter in the both observations and our simulations can in parts be attributed to different stellar populations, projection effects and global characteristics of the clouds. Furthermore, our findings of an increasing $\deficit$ with larger portions of clouds considered is in good agreement with the results for OMC~1 by \citet{Goicoechea2015}.

Also the global evolution of $\deficitl$ reported in our work (Fig.~\ref{fig:global_deficit_vs_tsf}, scales of $\lesssim 100$~pc) is in good agreement with observations. \citet{Suzuki2021} find values ranging between $10^{-3}$ and $10^{-2}$ in the star-forming region RCW~36. In particular, the authors observe a decreasing $\deficitl$ ratio for increasing $L_\mathrm{FIR}$, which is in excellent agreement with the evolution for $\deficitl$ found by us: in our simulations $\deficitl$ decreases with time, while $L_\mathrm{FIR}$ increases (Fig.~\ref{fig:global_deficit_vs_tsf}), which corresponds to a decrease in $\deficitl$ with increasing $L_\mathrm{FIR}$ as well.

The relation between $L_\mathrm{[CII]}$ and $L_\mathrm{FIR}$ has also been analysed for extragalactic observations. For M51, \citet{Pineda2018} find $\deficitl \simeq 0.003$ in the centre of the galaxy, and somewhat larger values in the inter-arm regions, generally matching the values we present here. Observations of samples of galaxies \citep{Diaz2017, Smith2017, Herrera2018} show $\deficitl$ values between a few $10^{-4}$ and $10^{-2}$ as well. Specifically, the observations also show a decrease of $\deficitl$ with increasing $L_\mathrm{FIR}$. As noted before, this qualitative behaviour matches well our findings. \citet{Sutter2021} studied 18 Local Volume galaxies and found that HII regions play a significant role in the overall [CII] deficit in their observed sample, in agreement with our findings. Although the authors state that thermalisation, that is collisional de-excitation, of C$^+$ plays a significant role in setting the $\deficitl$ ratio, they also argue that it cannot be the only effect. In the extragalactic context, the recent galaxy-scale hydrochemical models of \citet{Liang2024} find that the origin of the [CII] deficit can be associated with lower metallicities, an effect not studied here. However, the astrochemical models of \citet{Bisbas2024} using more realistic C/O ratios in low-metallicity environments, find only a weak correlation with the [CII] deficit, in agreement with the high-$z$ observations of \citet{Harikane2020}.

Furthermore, numerical simulations \citep{Smith2017, Bisbas2022} fit the relation between $\deficitl$ and the star formation rate surface density $\Sigma_{\mathrm{SFR}}$ by means of a power-law with an exponent of $-1/4.7$. In our work we refrained from investigating this correlation, as on cloud scales -- contrary to the galactic scales considered by the aforementioned authors -- the momentary SFR can vary significantly. However, we do find $\deficitl$ to be correlated with time (and thus increasing stellar mass) and in particular with the stellar luminosity (see Fig.~\ref{fig:global_deficit_vs_star_luminosity}). Specifically, we also find that the dependence of $\deficitl$ on $\lstartot$ can be well described by a power-law relation (Eq.~\ref{eq:bestfit}).

\section{Conclusions}\label{sec:conclusions}

In this paper, we present an analysis of simulations of four star-forming MCs within the SILCC-Zoom project \citep{Seifried2017,Haid2019}, including stellar feedback and an on-the-fly chemical network.
In particular, we analyse the [CII] 158~$\mu$m line and the FIR dust continuum emission. The aim of this paper is to investigate the origin and evolution of the so-called [CII] deficit, which is the observed drop in the $\deficitl$ ratio with increasing SFR and infrared luminosity. 

We performed two analyses, considering (i) spatially resolved emission inside MCs and HII regions and (ii) averages over the entire cloud scale. First, concerning the resolved [CII] and FIR emission in and around individual HII regions, we report the following results:

\begin{itemize}
    \item Due to the heating of dust by stellar radiation, $I_\mathrm{FIR}$ is initially high in the vicinity of newly born stars, and then moderately decreases over time in the centre of the HII regions as the HII regions get evacuated and the gas gets compressed into dense and cool shells.   
    \item On the other hand, $I_\mathrm{CII}$ only experiences a moderate increase after the onset of star formation and then quickly drops over time within the HII region. In addition to the intrinsic dilution of the gas inside the HII regions, we identify the further photo-ionisation of C$^+$ into C$^2+$ as a major contributor to this decrease, reducing the C$^+$ content to a few percent of its original value.     
    \item In combination, the two effects lead to a strong drop in $\deficit$ in the vicinity of newly born stars, that is, in the HII regions surrounding them. Specifically, we find that $\deficit$ decreases from values of 10$^{-3}$--10$^{-2}$ at scales above 10~pc to values of around 10$^{-6}$--10$^{-4}$ at scales below 2~pc.   
    \item Projection effects can significantly change the radial profile of $I_\mathrm{[CII]}$ and $I_\mathrm{FIR}$, and their ratio. Depending on the geometry of the cloud and on the viewing angle, an HII region may not be visible at all. Conversely, foreground material can cause apparent bubbles that do not host any stars.
\end{itemize}

Considering the evolution of total luminosities and the origin of the [CII] deficit on larger scales ($\sim100$~pc, corresponding to extragalactic observations), we report the following findings: 
\begin{itemize}
\item The $\deficitl$ ratio decreases from values of $\gtrsim$10$^{-2}$ in MCs without star formation to values of around $\sim10^{-3}$ in MCs with star formation within $\sim$1~Myr.
\item We attribute this decrease and thus the origin of the [CII] deficit to two main contributors: (i) the saturation of the [CII] line and (ii) the conversion of C$^+$ into C$^{2+}$ by stellar radiation.
\item The aforementioned drop in the $\deficitl$ ratio can be divided into two phases after the onset of star formation:
during the early evolution of HII regions, the saturation of [CII] and the further ionisation of C$^+$ limit the increase in $L_\mathrm{[CII]}$ as star formation progresses, while $L_\mathrm{FIR}$ increases rapidly, leading to an initial decline of $\deficitl$.   
\item In more evolved HII regions, $L_\mathrm{CII}$ stagnates and even partially drops over time due to the aforementioned secondary ionisation of C$^+$ and the saturation of [CII]. $L_\mathrm{FIR}$ also stagnates as the gas gets pushed into the cooler shells surrounding the HII region. In combination, this keeps the global $\deficitl$ ratio at low values of $\sim10^{-3}$.
\end{itemize}

Overall, our work shows that the evolution of the [CII] deficit is linked to the various evolutionary stages of MCs and their embedded star formation process. During this process different mechanisms contribute to varying degrees to the [CII] deficit. Our work can thus aid in interpreting observations of [CII] emission from scales of individual HII regions to galactic scales.

\begin{acknowledgements}
The authors thank the anonymous referee for the constructive report, which helped to increase the quality of the paper. The authors also thank C. Pabst for providing the observational data reused here in Fig.~\ref{fig:cii_vs_fir_colorbar_distance}.
 SE, DS, and SW thank the Deutsche Forschungsgemeinschaft (DFG) for funding through the SFB~956 ``Conditions and Impact of Star Formation'' (projects C5 and C6) and SFB~1601 ``Habitats of massive stars across cosmic time'' (sub-projects B1, B4 and B6). Furthermore, the project is receiving funding from the programme “Profilbildung 2020", an initiative of the Ministry of Culture and Science of the State of Northrhine Westphalia. TGB acknowledges support from DFG Grant No. 424563772, and the Leading Innovation and Entrepreneurship Team of Zhejiang Province of China (Grant No. 2023R01008). The software used in this work was in part developed by the DOE NNSA-ASC OASCR Flash Center at the University of Chicago.
 We particularly thank the Regional Computing Center Cologne of the University of Cologne for providing the computational facilities for this project by hosting our computing cluster ``Odin''.
\end{acknowledgements}

\bibliographystyle{aa}
\bibliography{biblio}

\begin{appendix}

\section{The radiation field below 13.6 eV}
\label{sec:G0}

As stated in Section~\ref{sec:silcc_zoom_simulations}, we do not include the effect of the stars on the radiation field at energies below 13.6~eV in the simulations. Doing so would have the same effect as locally increasing the $G_0$ factor for the ISRF. In an effort to get a rough estimation of how large this increase of $G_0$ would be, we present a here a first-order estimate. First, given that for each massive star we know its luminosity and surface temperature $T_\star$, we calculate the total luminosity of all stars in a MC in the range between 6 and 13.6~eV, $L_{\star, \textrm{ISRF}}$ \citep[corresponding to the photon energy range of the ISRF,][]{Draine1978}. Next, we make the assumption that all stars in a given MC are located at the same position. With this assumption, we can now convert $L_{\star, \textrm{ISRF}}$ into an effective $G_\textrm{0, eff}$, which would be present at an assumed distance $d$ from the stars (via geometrical attenuation $\propto d^{-2}$). Given the typical extent of the clouds and HII regions presented in this work, we choose $d$ = 10, 20 and 30~pc. We note that in this first-order estimate we neglect attenuation by dust.

\begin{figure}
            \centering
            \includegraphics[width=0.99\columnwidth]{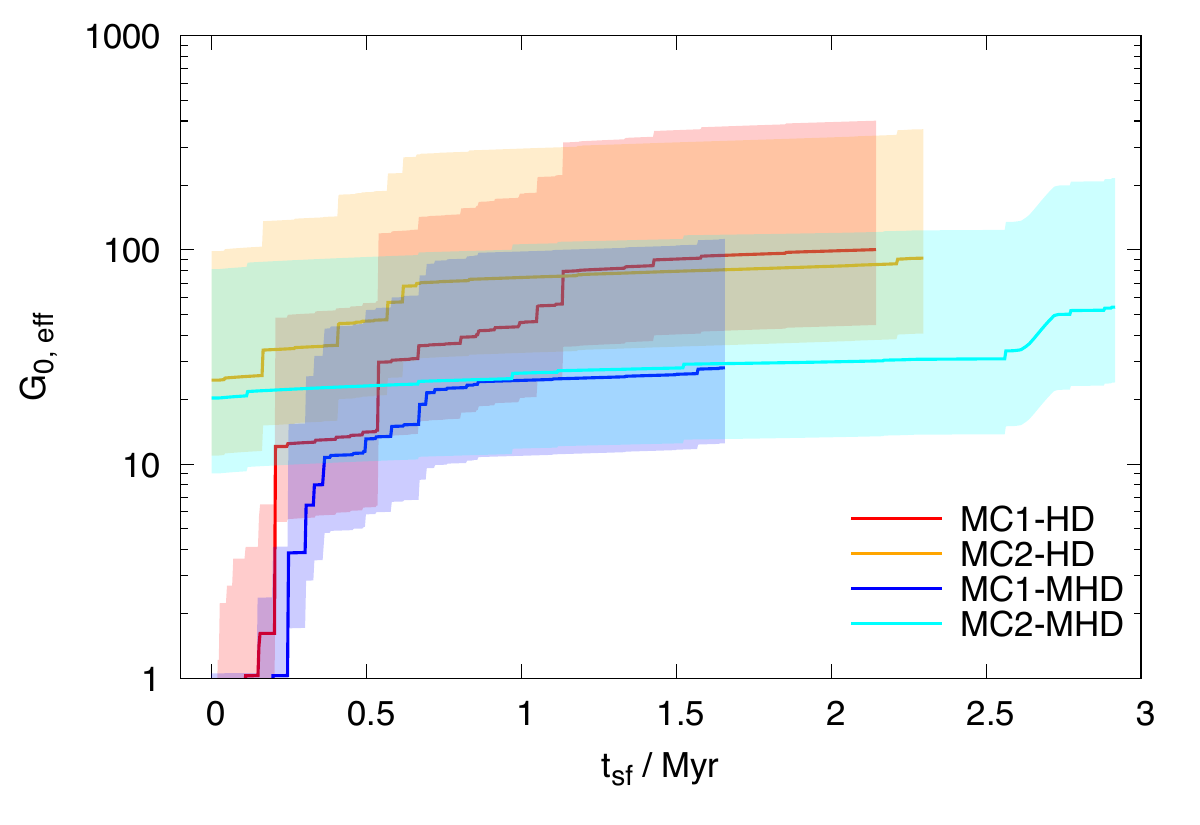}
            \caption{Effective $G_\textrm{0, eff}$ created by the massive stars in the different clouds (see text). The solid line represents $G_\textrm{0, eff}$ at an assumed distance of 20~pc, the shaded area the range for distances between 10~pc and 30~pc. Typical values are of the order of a few 10 - 100.}
            \label{fig:G0}
        \end{figure}

We plot the resulting $G_\textrm{0, eff}$ for all four clouds considered in this work in Fig.~\ref{fig:G0}. As can be seen, the effective $G_\textrm{0, eff}$ created by the stars increases over time from a few 10 to $\sim$100. We again stress that these values present only a very rough estimate, the actual increase in $G_0$ might vary significantly depending on the chosen location in the cloud. Nonetheless, the substantial increase of $G_0$ with respect to our fiducial value of 1.7 could have some effect on the temperature profile of the HII regions. In particular in the surrounding shells, where the ionising radiation is already extincted -- and thus also the Cloudy post-processing step described in Section~\ref{sec:post_processing} is not applied -- we do not use the $G_\textrm{0, eff}$ produced by the nearby stars. This could slightly alter the temperature as well as the CO and thus C$^+$ abundance in the PDR region. However, in the volume-wise larger interior of the HII regions, the effect would presumable be only of minor importance as here the heating of the gas would mainly be due to the ionising radiation, which is taken into account in the simulations. Hence, we expect that taking into account radiation below 13.6~eV would have only a moderate effect in a spatially very confined region.

\section{Monte-Carlo resolution study}
\label{sec:resstudy}

        \begin{figure}
            \centering
            \includegraphics[width=0.99\columnwidth]{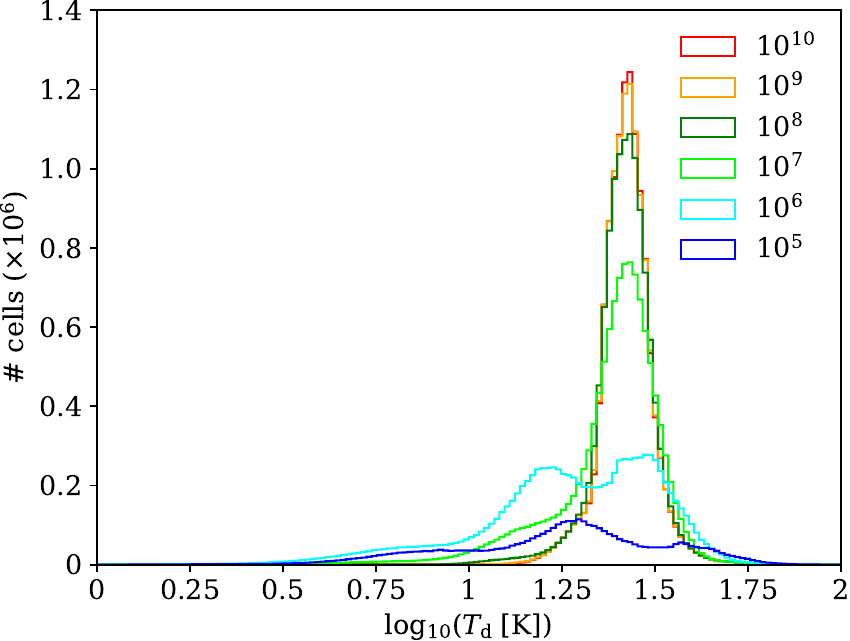}
            \caption{Dust temperature distribution calculated with RADMC-3D of MC2-HD at $\tevol = 4$ Myr for different numbers of photon packages. Differences between the distribution for $10^9$ and $10^{10}$ photon packages are less than 1 per cent for all dust temperature bins, and therefore we chose to use $10^{10}$ photon packages for all dust temperature calculations, as the computational costs for this highest value are still manageable.}
            \label{fig:convergence_dust_temperature}
        \end{figure}
        
We perform a convergence study to determine a sufficient number of RADMC-3D photon packages for the calculation of $T_\mathrm{d}$. To do so, we consider MC2-HD at $\tevol = 4$ Myr, as this is the snapshot which contains the largest number of stars. As RADMC-3D splits up the available photon packages to the sources (i.e. stars), the number of packages per star is the lowest for this snapshot. Hence, convergence for this snapshot should also imply convergence for all other snapshots. In Fig.~\ref{fig:convergence_dust_temperature} we show the distribution of $T_\mathrm{d}$ (calculated by RADMC-3D) for different numbers of photon packages. We note that with a low number of photon packages ($10^5 - 10^6$) the distribution is clearly not yet converged, whereas the distributions for $10^9$ and $10^{10}$ photon packages are almost identical (they vary for less than 1 per cent). We thus choose to use $10^{10}$ photon packages in the Monte Carlo dust temperature calculation. 

\end{appendix}

\end{document}